\documentclass{jfm}
\usepackage{graphicx}

\usepackage{color}

\usepackage{psfrag}
\usepackage{amsmath}
\usepackage{amssymb}
\usepackage{mathrsfs}
\usepackage{subfigure}
\usepackage{amsfonts}
\usepackage{bm}
\usepackage{amssymb}
\usepackage[section]{placeins}
\usepackage{epstopdf}
\usepackage{float}
\usepackage{breqn}
\usepackage{tabu}
\usepackage{booktabs} 
\usepackage{natbib}

\newcommand*{\beq}{\begin {equation} }
\newcommand*{\eeq}{\end {equation} }
\newcommand*{\beqs}{\begin {equation*} }
\newcommand*{\eeqs}{\end {equation*} }

\title[On the role of actuation for the control of streaky structures in boundary layers]
{On the role of actuation for the control of streaky structures in boundary layers}

\author[K. Sasaki, P. Morra, A. V. G. Cavalieri, A. Hanifi, D. Henningson]%
{Kenzo Sasaki$^{1,2}$,\ns
Pierluigi Morra$^{2}$\break
Andr\'e V. G. Cavalieri$^{1}$,\ns
Ardeshir Hanifi$^2$\break
Dan S. Henningson$^{2}$}
\affiliation{$^1$Instituto Tecnol\'ogico de Aeron\'autica, S\~ao Jos\'e dos Campos, Brazil\\[\affilskip]
$^2$KTH Royal Institute of Technology, Linn\'e FLOW Centre, SE-10044, Stockholm, Sweden
}

\volume{???}
\pagerange{???--???}
\date{?; revised ?; accepted ?. - To be entered by editorial office}

\begin{document}

\maketitle

\begin{abstract}
This work deals with the closed-loop control of streaky structures induced by free-stream turbulence (FST) in a zero-pressure gradient, transitional boundary layer, by means of localized sensors and actuators. A linear quadratic gaussian regulator is considered along with a system identification technique to build reduced-order models for control. Three actuators are developed with different spatial supports, corresponding to a baseline shape with only vertical forcing, and to two other shapes obtained by different optimization procedures. A computationally efficient method is derived to obtain an actuator which aims to induce the exact structures which are inside the boundary layer, given in terms of their first spectral proper orthogonal decomposition (SPOD) mode, and an actuator that maximizes the energy of induced downstream structures. Two free-stream turbulence levels were evaluated, corresponding to 3.0\% and 3.5\%, and closed-loop control is applied in large-eddy simulations of transitional boundary layers. All three actuators lead to significant delays in the transition to turbulence and were shown to be robust to mild variations in the free-stream turbulence levels. Differences are understood in terms of the SPOD of actuation and FST-induced fields along with the causality of the control scheme when a cancellation of disturbances is considered. The actuator optimized to generate the leading downstream SPOD mode, representing the streaks in the open-loop flow, leads to the highest transition delay, which can be understood due to its capability of closely cancelling structures in the boundary layer. However, it is shown that even with the actuator located downstream of the input measurement it may become impossible to cancel incoming disturbances in a causal way, depending on the wall-normal position of the output and on the actuator considered, which limits sensor and actuator placement capable of good closed-loop performance.   
\end{abstract}

\section{Introduction}
\label{intro}

The control of transitional and turbulent boundary layers in wall-bounded flows could potentially lead to high benefits in terms of energy saving, owing to the fact that up to 50\% of drag and therefore fuel consumptions of modern aircraft is to overcome the skin friction due to turbulent boundary layers \citep{schrauf2005status}, a ratio which is expected to increase due to the growing wing aspect ratio of new generation of aircraft and corresponding reduction in wave and induced drag components. Any reduction on the skin friction will result in significant savings for the operational cost of such aircraft. 

\subsection{Transition to turbulence}

In the classical route to turbulence, which occurs in a low-perturbation scenario, a laminar boundary-layer solution becomes unstable to infinitesimal perturbations, which will grow exponentially in the form of two-dimensional Tollmien-Schlichting (TS) waves. When a critical amplitude for such fluctuations is reached, nonlinear interactions start to occur, which will eventually lead to three-dimensionality and breakdown to turbulence, a process which is thoroughly described in the review of \cite{kachanov1994physical}.

However, if the zero-pressure gradient laminar boundary layer is subject to levels of free-stream turbulence higher than $\approx 1 \%$ the transition to turbulence will occur via a different mechanism, which ``bypasses" the classical Tollmien-Schlichting case \citep{matsubara2001disturbance}. Such behaviour is explained by the non-normality of the Orr-Sommerfeld/Squire operator which describes the flow dynamics, which is associated to non-orthogonal eigenmodes \citep{reddy1993energy,schmid2012stability}. Such non-orthogonality may lead to strong transient amplifications, which may occur even if the flow is stable. In the case of boundary layers, the upstream perturbations which undergo the highest transient amplifications take the form of streamwise elongated structures with comparably narrow spanwise scales. Such streaky structures are sometimes referred to as the Klebanoff mode, referencing to the experiments of \cite{klebanoff1971effect}; more recent experiments have also identified such structures for different levels of free-stream turbulence \citep{westin1994experiments,matsubara2001disturbance}. In the works of \cite{andersson1999optimal} and \cite{luchini2000reynolds} it has been shown that the free-stream turbulence generates streaky structures matching those generated by the optimal perturbation, calculated from a transient growth analysis.

The physical origin of these streaks  may be explained by the lift-up effect \citep{ellingsen1975stability,landahl1980note}, where wall-normal velocity disturbances cause the movement of fluid across the boundary layer; low-speed fluid is pushed away from the wall and high-speed fluid is pushed towards it. This movement creates quasi-periodic low and high-speed streaks which will grow linearly in the streamwise direction. With the growing intensity of the streaks, they become susceptible to higher-frequency secondary instabilities \citep{brandt2002transition,brandt2002weakly}, which will develop in turbulent spots, localized regions of chaotic motion \citep{brandt2004transition,schlatter2008streak}. When these spots merge, they will lead to a fully developed turbulent boundary layer.

\subsection{Control}

The high-dimensionality and inherent nonlinearity of the Navier-Stokes equations cause the computational requirements both of the simulated system and online actuation calculation to rapidly become intractable with the size of the calculation domain. The usual strategy to overcome these difficulties consists in the ``reduce-then-design'' approach \citep{semeraro2013riccati}, where the control-laws are designed off-line in a reduced-order model and tested \textit{a posteriori} in the full nonlinear system, either a simulation or experiment \citep{bagheri2009input,belson2013feedback,semeraro2013transition,semeraro2011feedback}
.

Once the reduced-order model is available, a common strategy for control of boundary layer transition is to place the actuation in a region where the amplitude of the perturbations is small and to account for the convective nature of the flow via a feedforward scheme, where the actuator is placed downstream of the input and upstream of the objective position. The control action is then decided by means of measuring the input and acting to minimize a given quantity at the objective position. This can be accomplished using static compensators, such as the Linear Quadratic Gaussian (LQG) regulator \citep{barbagallo_sipp_schmid_2009,barbagallo2011input,juillet2014experimental,schmid2016linear,fabbiane2017energy}.

The previously cited works deal with the control of the transition induced by modal instabilities, such as TS waves. The control of non-modal structures is more rare and applications may be found in the works of \cite{jacobson1998active,hanson2010transient,papadakis2016closed,bade2016reactive}, all of which deal with isolated streaks. This implies that the streaks are generated inside the boundary layer, either via roughness elements or via the inclusion of pairs of modes in a numerical simulation.

In less artificial studies, \cite{lundell2007reactive} and \cite{monokrousos2008dns} used blowing and suction at the wall and wall-shear stress measurements combined with feedforward control to delay the transition induced by free-stream turbulence, which inherently considers a much greater number of upstream modes. However, Lundell tuned the control effort for one specific configuration and Monokrousos \textit{et al.} used spatially extended actuators with many degrees of freedom which would be prohibitive for practical implementations. \cite{lundell2009feedback} demonstrated the drawbacks of currently available actuators and suggested they pose a considerable limitation for the control of streaky structures in flow control applications (see the review of \cite{cattafesta2011actuators} for an overview of actuators for flow control applications).

The difficulty in the control of bypass transtion is that, differently from the Tollmien-Schlichting case, which corresponds to a definite modal instability, a family of streaks may be generated inside the boundary layer. Even though the resulting structure will correspond to the one generated by the optimal perturbation, as shown in \cite{luchini2000reynolds}, its precise shape will be different depending on where it is generated. This poses a challenge to the actuator which in practice has to be located in an specific position. 

\subsection{Contribution of the present work}

The present study tackles the mitigation of unsteady streaks, generated by means of free-stream turbulence, which penetrates the boundary layer via the receptivity mechanism \citep{brandt2004transition}. We assess the role of actuation by considering different strategies for the design of the resulting forcing, which gives insight into the physics behind the active control of streaks. Such strategies are useful for the design and evaluation of actuators for the active control of streaky structures.

The paper is organized as follows. Section \ref{flowestcont} introduces the flow configuration control and estimation methods. Spectral proper orthogonal decomposition (SPOD) is applied to the open-loop data in Section \ref{spodfortheestimation}, the result being compared to the optimal perturbation. The methods for the design of actuators are given in Section \ref{actuatorstobeconsidered} with the results and discussion in Sections \ref{resultsfortransitiondelaymainly} and \ref{understandingtheresults}, respectively; finally, conclusions are drawn in Section \ref{concludingthepaper}. The appendix presents the specifics of the SPOD calculation and a detailed description of the adjoint optimization methods considered in the design of the forcings.

\section{Flow configuration, control methods and estimation tools}
\label{flowestcont}

\subsection{Flow configuration}

The incompressible Navier-Stokes equations model the flow,

\begin{equation}
\label{momentumequation}
\frac{\partial \mathbf{q}}{\partial t}+(\mathbf{q} \cdot \nabla)\mathbf{q}=-\nabla p
+\frac{1}{Re}\nabla^2\mathbf{q}+\lambda_{fringe} (x_1)\mathbf{q}+\mathbf{f}
\end{equation}

\begin{equation}
\label{continuityequation}
\nabla \cdot \mathbf{q}=0
\end{equation}

\noindent where $\mathbf{q}(\mathbf{x},t)=(u(\mathbf{x},t),v(\mathbf{x},t),w(\mathbf{x},t))$ and $p(\mathbf{x},t)$ are the velocity and pressure, respectively, at each time step $t$ and position $\mathbf{x}=(x_1,x_2,x_3)$, taken in the cartesian coordinates.

The same flow configuration of the parallel investigation of \cite{morra2019_inpress} will be considered here. A plate of semi-infinite length lies in the $x_1x_3$ plane, where no-slip conditions are enforced at $x_2=0$. The control action is analysed via large-eddy Simulations (LES) with the pseudo-spectral code SIMSON \citep{chevalier2007simson}, which gives a high numerical accuracy. The flow is periodic along the spanwise direction and a fringe forcing, given as $\lambda_{fringe}(x_1)$, is introduced in the last 20\% of the domain to ensure periodicity also along the streamwise direction. Spatial coordinates and velocities are non-dimensionalized using the displacement thickness $\delta^{*}$ in the entrance of the domain and the free-stream velocity $U_{\infty}$, respectively. The resulting Reynolds number, defined as $Re=\delta^{*} U_{\infty}/\nu$, where $\nu$ is the kinematic viscosity, is 300. The computational domain for the 3D simulation is of $[0,4000] \times [0,30] \times [-25,25]$ in the $x_1$, $x_2$ and $x_3$ directions, with $N_{x_1}=1024$ and $N_{x_3}=108$ Fourier modes discretizing the $x_1x_3$ plane and $N_{x_2}=121$ Chebyshev polynomials in the vertical direction. 

A volume forcing $\mathbf{f}$ is used to perform the control action, and its spatial shape will be obtained by three different methods, which will be introduced in section \ref{actuatorstobeconsidered}. 

At the fringe region a number of modes from the continuous branch of the Orr-Sommerfeld Squire operators (which will be referred to as OSS modes) is forced. The prescribed perturbation takes the form of 

\begin{equation}
\label{euqaiontaforossmodes}
\mathbf{q}^{\prime}_{FST}=\sum_{\alpha}\sum_{\beta}\sum_{\omega}\phi(\alpha,\beta,\omega)\mathbf{q}^{\prime \star}(x_2,\alpha,\beta,\omega)e^{i(\alpha x_1+\beta x_3 -\omega t)}
\end{equation}

\noindent where $\mathbf{q}^{\prime}=(u^{\prime},v^{\prime},w^{\prime})$, the prime indicates a fluctuation and $q^{\prime \star}$ represents the eigensolution of the Orr-Sommerfeld Squire eigenvalue problem for the velocity fluctuations for a parallel flow, $\alpha$, $\beta$ and $\omega$ are the streamwise and spanwise wavenumber and the angular frequency, respectively. For further details concerning the method, the reader is referred to the work of \cite{brandt2004transition}. A number of 200 modes, with an integral length scale of $L=7.5 \delta^*$ and turbulent intensity of 3.0\% or 3.5\% will be considered in this work. The characteristic spectrum of the free-stream turbulence seeks to represent the von Karman spectrum and is the same as in \cite{brandt2004transition} and also used in \cite{morra2019_inpress} to produce homogeneous isotropic turbulence. For further details the reader is referred to the previously cited works.

\begin{figure}
\centering
\includegraphics[width=0.80\textwidth]{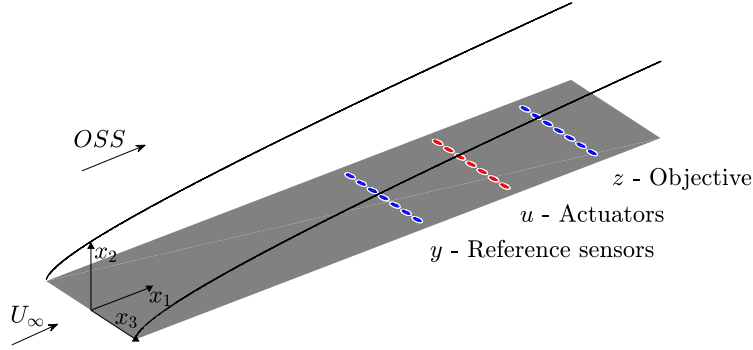}
\caption{Scheme for the 3D simulation of the flat plate considered. The blue and red circles represent the input sensors and actuators, respectively.}
\label{schemefortheboundarylayersimulation}
\end{figure}

A localized measurement of the streamwise skin friction is used to define the inputs given by sensors $\mathbf{y}(t,x_3)$, and downstream objective, $\mathbf{
z}(t,x_3)$. Three rows of 36 equispaced independent objects are placed with a transverse separation of $\Delta x_3=1.4$, which is adequate to resolve the spanwise wavenumber content of the fluctuations considered here. Measurements are taken at $x_{1_y}=250$ and $x_{1_z}=400$, defining input and objective, respectively. Actuation is performed at $x_{1_u}=325$. Alternatively, streamwise positions will sometimes be referred to by the local Reynolds number based on $x_1$, $Re_x$. For sensor, actuation and objective positions, $Re_x$ is equal to 105000, 127000 and 150000, respectively. Figure \ref{schemefortheboundarylayersimulation} presents a scheme for the current simulation and coordinates considered in this paper. 

\subsection{Estimation tools and control methods}

For the development of the control law, the same approach as per \cite{morra2019_inpress} will be followed with the construction of a linear quadratic gaussian (LQG) regulator \citep{bagheri2009input2,fabbiane2015role,sasaki2018wave}, using the eigensystem-realization-algorithm (ERA) \citep{juang1985eigensystem,ma2011reduced} to supply a state-space representation of tractable dimension for the design of the LQG controller. 

The actuation is computed from a convolution of the measurements $\mathbf{y}(t,x_3)$ with a kernel $\mathbf{k}(t,x_3)$. The spanwise direction is discretized considering the position of the localized sensors and actuators, such that each actuator will behave as

\begin{equation}
\label{discretizedversionofthe3dkernel}
u_l(n)=\int_0^t\sum_{m=0}^{N}k_m(t-\tau)y_{l-m}(t-\tau) \mathrm{d}\tau,
\end{equation}

\noindent where the index $l$ refers to each actuator and sensor, such that all the sensor measurements are considered in the computation of the actuation signal of each actuator.

The design of the LQG regulator involves the solution of two Riccati equations, one for the Kalman gain and one for the actual control kernel. Such calculation requires a state-space description of the problem, which is given in terms of a matrix $\mathbf{A}$, describing the system dynamics, matrices $\mathbf{B}$ and $\mathbf{M_d}$ which describe the effect of actuation of the disturbance and matrices $\mathbf{C_y}$ and $\mathbf{C_z}$ determining the actual measurements,

\begin{equation}\label{eq:LQGsys00}
\begin{aligned}
\dot{\mathbf{q}} &= \mathbf{A} \mathbf{q} + \mathbf{B} \mathbf{u} + \mathbf{M}_d \mathbf{d},\\
\mathbf{y} &= \mathbf{C}_y \mathbf{q} + \mathbf{n} ,\\
\mathbf{z} &= \mathbf{C}_z \mathbf{q},\\
\end{aligned}
\end{equation}

\noindent white noise $\mathbf{n}$ is also assumed to be present in the measurement sensor. The numerous degrees of freedom of typical fluid mechanics problems require the usage of a reduced-order model for the description of equation \eqref{eq:LQGsys00}. As in previous works by this group \citep{sasaki2018wave}, the eigensystem realization algorithm (ERA) \citep{juang1985eigensystem} was chosen for this task. ERA involves the singular value decomposition of a Hankel matrix, formed by the impulse responses of all the inputs of the system which, for this case, correspond to the disturbances $\mathbf{d}$ and actuation $\mathbf{u}$. For the details concerning the construction of the Hankel matrix, the reader is referred to \cite{sasaki2018wave}.

The difficulty here is that the considered disturbance is formed by a great number of OSS modes, which implies that the computation of each individual impulse response is not feasible computationally. Furthermore, such impulse responses are not available for the case of an experimental implementation. We therefore proceed by a somewhat different strategy, defining a new set of ``dummy'' measurements $\mathbf{y_d}$ which is placed upstream of the $\mathbf{y}$ and $\mathbf{z}$ measurements. Empirical transfer functions are then calculated between $\mathbf{y_d}$ and $\mathbf{y}$ or $\mathbf{z}$, following the procedure introduced in \cite{morra2019_inpress},

\begin{equation}\label{eq:d2yz}
\hat{G}_{y_{d}y}(\omega,\beta_k)=\frac{\hat{S}_{y_{d}y}(\omega,\beta_k)}{\hat{S}_{y_{d}y_{d}}(\omega,\beta_k)},\quad \quad
\hat{G}_{y_{d}z}(\omega,\beta_k)=\frac{\hat{S}_{y_{d}z}(\omega,\beta_k)}{\hat{S}_{y_{d}y_{d}}(\omega,\beta_k)}.
\end{equation}

\noindent where $\hat{S}_{y_{d}y_{d}}(\omega,\beta_k)$ and $\hat{S}_{y_{d}y}(\omega,\beta_k)$ or $\hat{S}_{y_{d}z}(\omega,\beta_k)$ are the auto and cross-spectra of the dummy measurement and the measurements, $\mathbf{y}$ and $\mathbf{z}$, and are calculated via ensemble averaging. The discrete spanwise wavenumbers $\beta_k$ are defined by considering each localized actuator as a discrete measurement at a given position $x_3$, $\beta_k=[-\beta_{max}/2 \quad  \beta_{max}/2]$, where $\beta_{max}=2\pi/(\Delta x_3)$.

Inverse Fourier transforming the quantities defined in equation \eqref{eq:d2yz},


\begin{equation}
\label{inversefouriertransform1}
g_{y_{d}z}(t,x_3)=\frac{1}{2\pi}\frac{1}{N_{\beta}}\int_{-\infty}^{\infty} \sum_{k=0}^{N-1} \hat{G}_{y_{d}z}(\omega,\beta_k) e^{i \beta_k x_3}e^{-i \omega t} \mathrm{d}\omega,
\end{equation}

\noindent where $N_{\beta}$ is the number of discrete transverse wavenumbers considered, provides empirically identified impulse responses which may be directly applied in the ERA method for the construction of ROMs for designing LQG. This procedure based only on the measured signal, which permits the definition of LQG controllers even for experimental implementations, was firstly introduced in \cite{morra2019_inpress} and the reader is directed for such work for further details.

Application of ERA for this problem results in a system with $N_{ERA}=387$ modes, which accurately reproduces the empirically identified transfer functions.

\begin{figure}
\begin{center}
\subfigure[]{\includegraphics[width=0.48\textwidth]{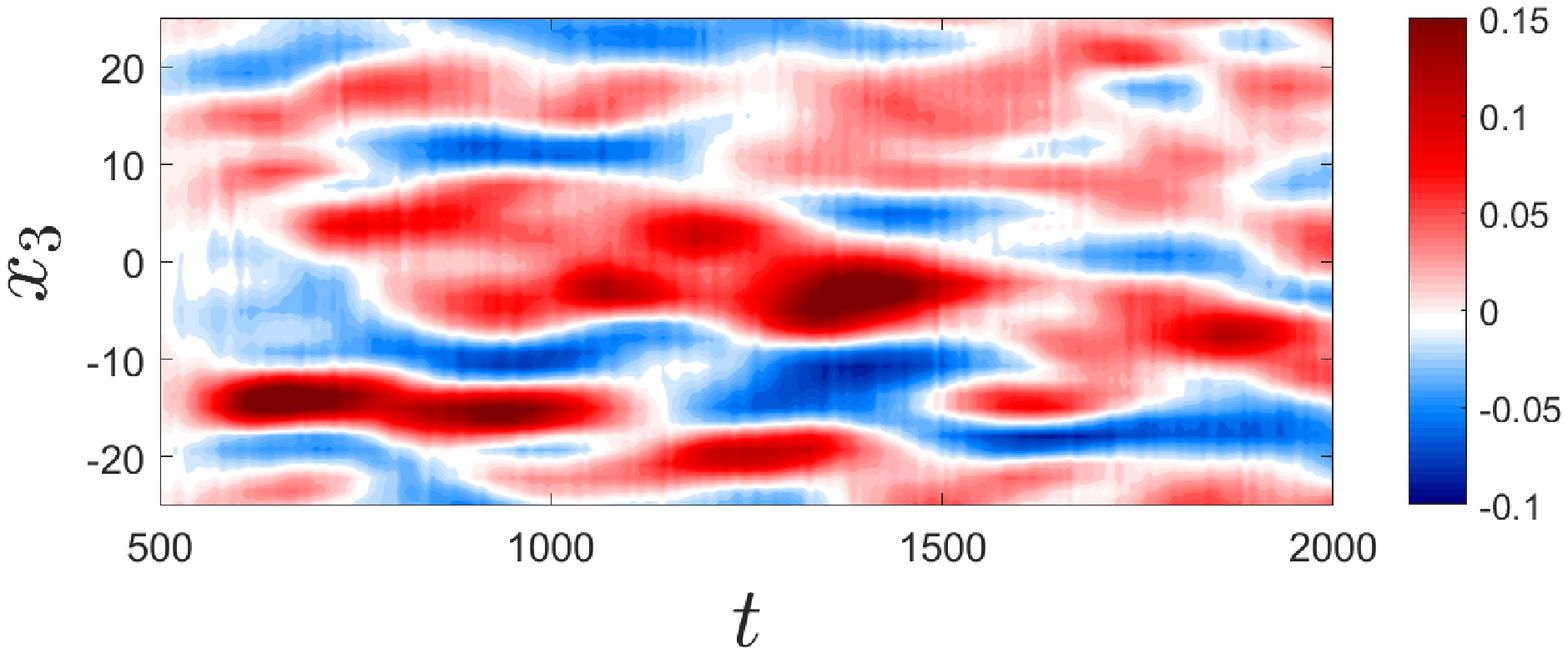}}
\subfigure[]{\includegraphics[width=0.48\textwidth]{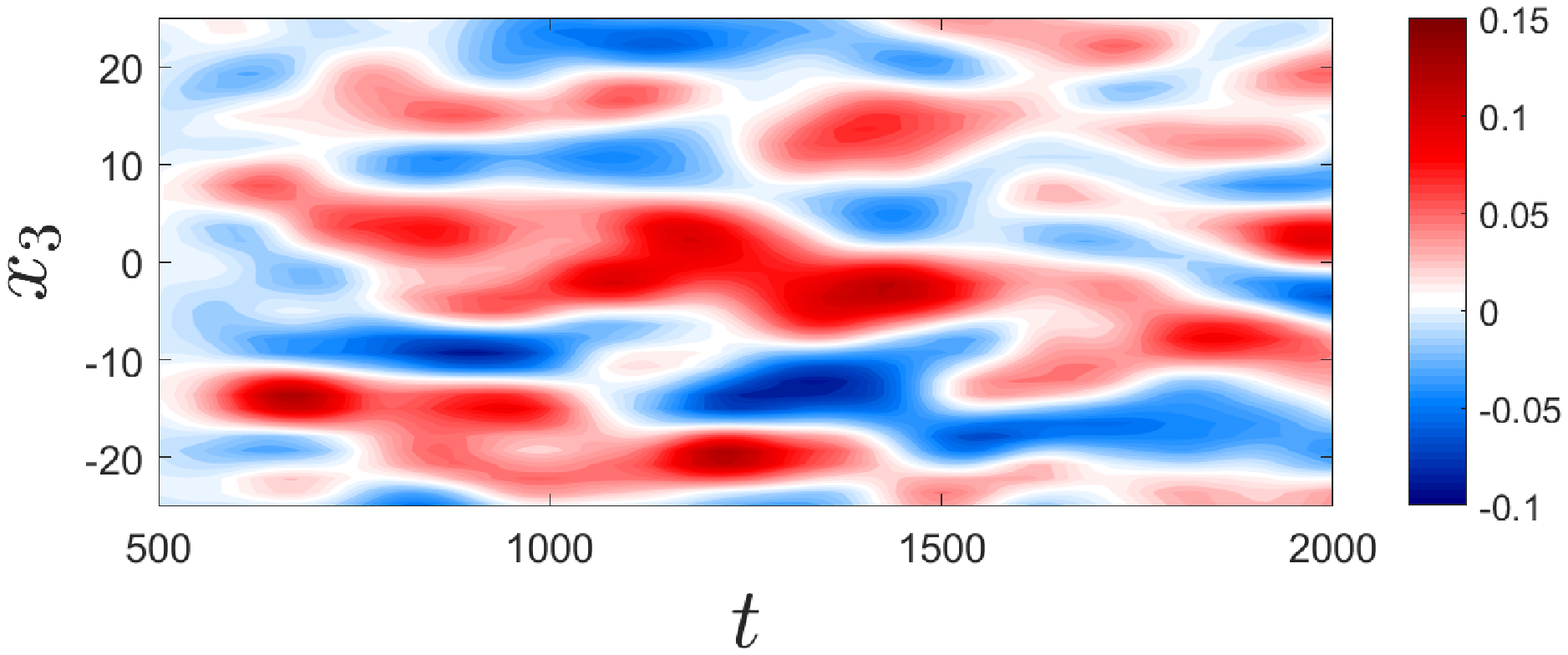}} \\
\end{center}
   \caption{Comparison between (a) LES field at $(x_1,x_2)=400$ and (b) its prediction from the empirical transfer function using wall-shear stress measurements at $(x_1,x_2)=(250,0)$.}
  \label{performanceofthereducedordermodels}
\end{figure}

Such empirically derived transfer functions may also be used to predict the time and spanwise behaviour of the $\mathbf{z}(t,x_3)$ measurement, at the objective position, from the $\mathbf{y}(t,x_3)$ measurement, when the actuator is not active in the system. The empirical transfer function is then

\begin{equation}
\hat{G}_{yz}(\omega,\beta_k)=\frac{\hat{S}_{yz}(\omega,\beta_k)}{\hat{S}_{yy}(\omega,\beta_k)}
\end{equation}

\noindent with $g_{yz}(t,x_3)$ resulting from the double inverse Fourier transform, as per equation \eqref{inversefouriertransform1}. The prediction is then taken as the double convolution of $\hat{g}_{yz}(\omega,\beta_k)$ with the measurements $\mathbf{y}(t,x_3)$. This procedure may be applied to any streamwise separated measurements. Figure \ref{performanceofthereducedordermodels} presents a sample of the prediction of $\mathbf{z}(t,x_3)$ from the measurements $\mathbf{y}(t,x_3)$ for the set-up considered in this paper, and validates the procedure. For more details concerning the application of the proposed methodology for the time-domain prediction of streaky structures induced by free-stream turbulence, the reader is referred to the work of \cite{morra2019_inpress}, which firstly introduced the method for this type of application.

\section{Spectral proper orthogonal decomposition applied to transitional streaks}
\label{spodfortheestimation}

In what follows, spectral proper orthogonal decomposition (SPOD) is applied to fluctuation data at the $(x_2,x_3)$ cross-stream plane at the streamwise objective position, $x_1=400$, without any control action taking place. SPOD has been used in previous studies \citep{picard2000pressure,cavalieri2013wavepackets,semeraro2016stochastic,towne2018spectral} with the objective of extracting the most energetic, and probable, structures in the flow, for each $(\omega,\beta_k)$ combination. Here the SPOD modes will be used to extract the dominant structure in the flow, to determine the best suited actuator for this application, and, finally, to evaluate how the closed-loop actuation is attenuating the streaks in the flow.

SPOD is applied to the velocity fluctuations such that they are optimal modes to represent the turbulent kinetic energy, where the modes are defined from the solution of the following integral equation,

\begin{equation}
\label{spodequation1}
\int \mathbf{R}(\mathbf{x},\mathbf{x'},\omega,\beta_k)\psi_j(\mathbf{x'},\omega,\beta_k)\mathrm{d}\mathbf{x'}=\lambda \psi_i(\mathbf{x},\omega)
\end{equation}

\noindent where $\psi$ will correspond to an eigenfunction (SPOD mode) with corresponding $\lambda$, eigenvalue, and $\mathbf{R}(\mathbf{x},\mathbf{x'},\omega,\beta_k)$ is the two-point cross spectral density, which is defined from the Fourier transform of the correlation tensor,

\begin{equation}
\label{definingthecrosspowerspectraltensor}
\mathbf{R}(\mathbf{x},\mathbf{x'},\omega,\beta_k)=\int_{-\infty}^{\infty}\mathbf{C}(\mathbf{x},\mathbf{x'},\tau,\beta_k)e^{i\omega \tau}\mathrm{d}\tau.
\end{equation}

\noindent The correlation tensor $\mathbf{C}$ is obtained by:

\begin{equation}
\label{correlationtensor}
\mathbf{C}(\mathbf{x},\mathbf{x'},\tau,\beta_k)=E[\mathbf{q}(\mathbf{x},t)\mathbf{q}^*(\mathbf{x}^\prime,t^\prime+\tau)],
\end{equation}

\noindent with $\mathbf{q}=(u,v,w)$, the three velocity components, and $E[.]$ the expectation operator, representing the expected value of a given realization of the flow.

Equation \eqref{spodequation1} may be replaced by an eigenvalue problem \citep{towne2018spectral} which reads:

\begin{equation}
\label{spodequation2}
\mathbf{H}(\omega,\beta_k)\psi(\omega,\beta_k)=\lambda(\omega,\beta_k)\psi(\omega,\beta_k)
\end{equation}

\noindent where the elements of $\mathbf{H}(\omega,\beta_k)$ are calculated via an ensemble averaging,

\begin{equation}
\label{spodequation3}
H_{ij}(\omega,\beta_k)=\langle \hat{\mathbf{q}}_i(\omega,\beta_k) \hat{\mathbf{q}}_j(\omega,\beta_k) \rangle,
\end{equation}

\noindent where $\hat{\mathbf{q}}=(\hat{u}(\omega,\beta_k),\hat{v}(\omega,\beta_k),\hat{w}(\omega,\beta_k))$. For a detailed description of SPOD, the reader is referred to the work of \cite{towne2018spectral}, in Appendix A, a brief description of the application of SPOD to data is provided.

\begin{figure}
\begin{center}
\subfigure[]{\includegraphics[width=0.8\textwidth]{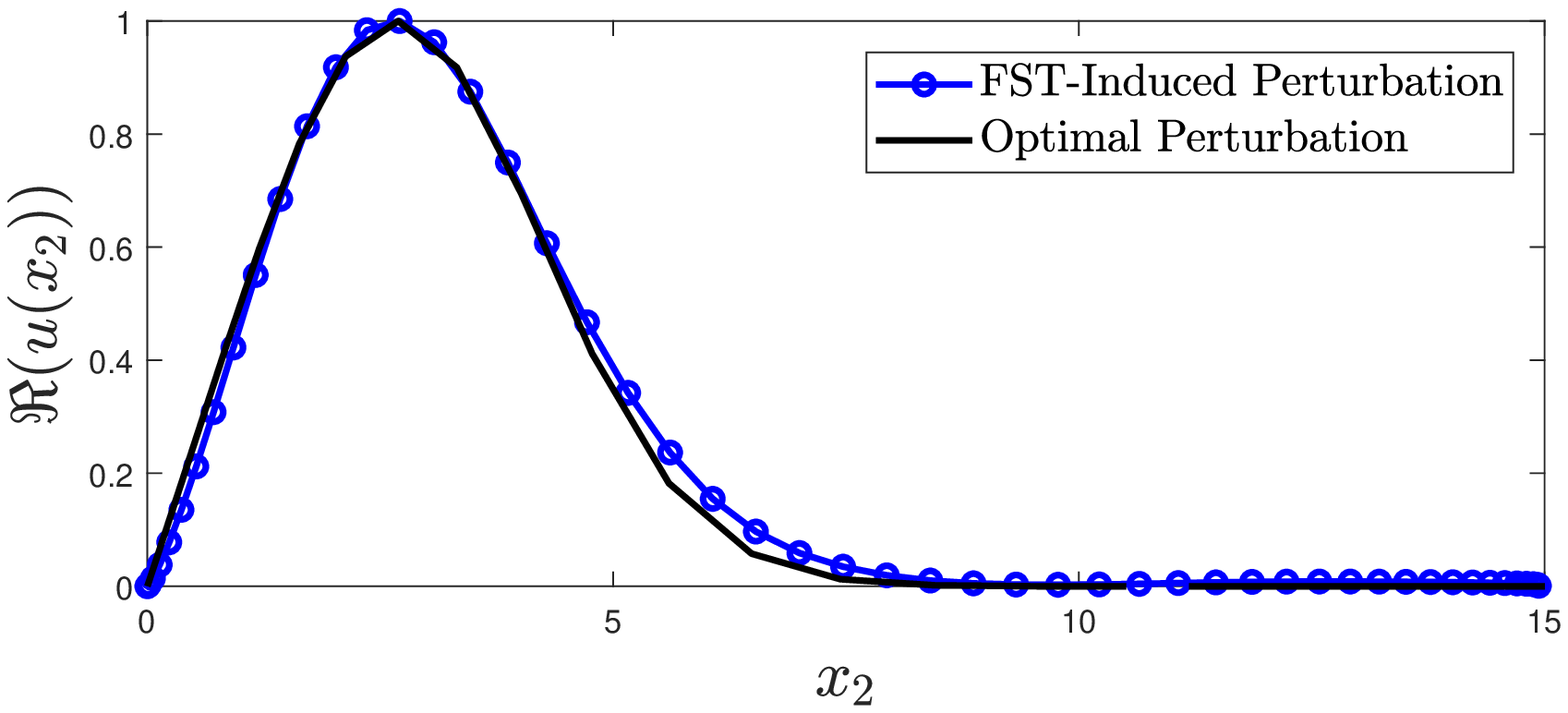}} \\
\subfigure[]{\includegraphics[width=0.8\textwidth]{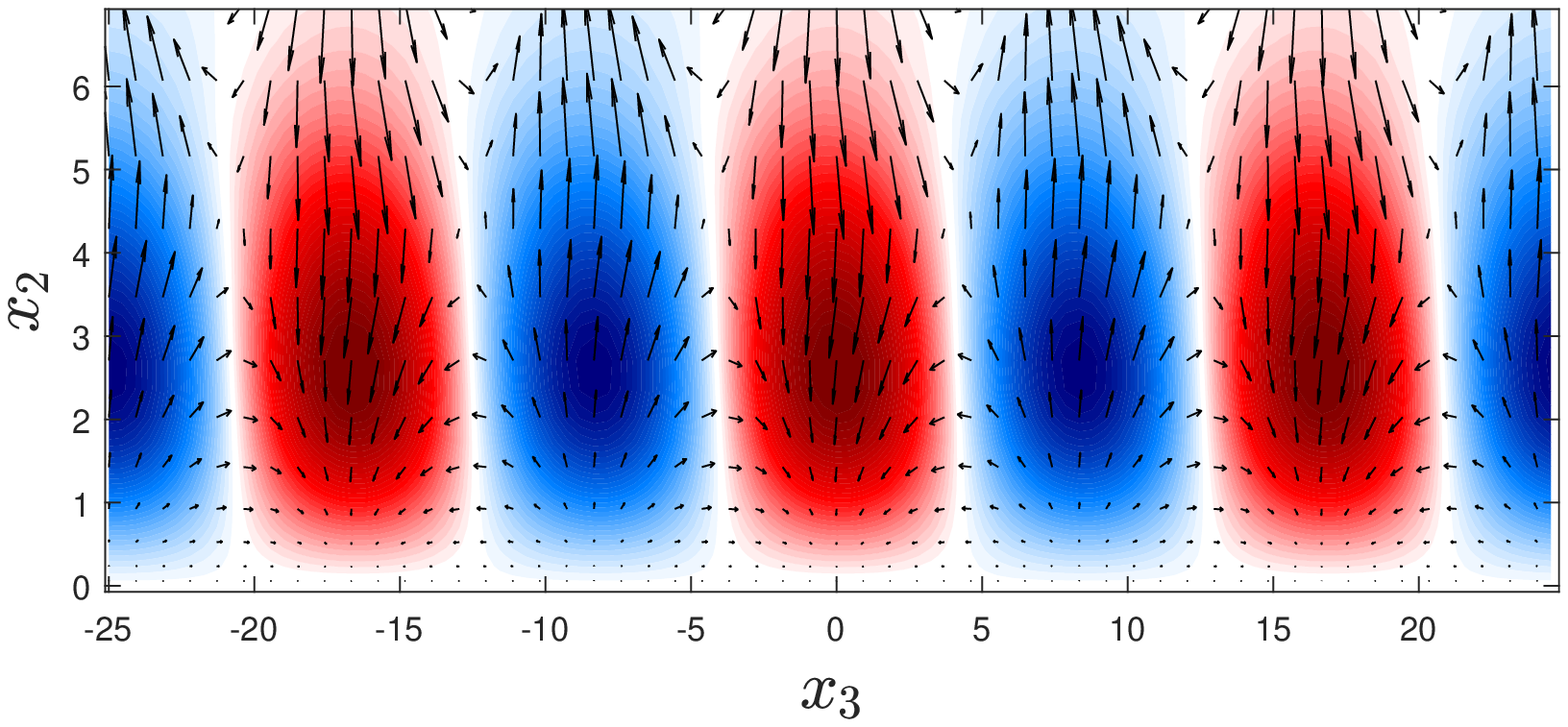}} \\
\end{center}
   \caption{Comparison of the SPOD modes in the wall-normal direction induced by the FST with the result of the optimal perturbation (a) and characteristic streaky behaviour observed in the streamwise velocity fluctuations (b), the pseudocolours vary between -1 and 1 and the arrows correspond to the velocity field induced by the wall-normal and spanwise velocity fluctuations. Modes were normalized to present unitary amplitude. Comparison for $Tu=3.0\%$}
  \label{podcomparisonandoptimalperturbation}
\end{figure}

The elements in equation \ref{spodequation3} are determined by means of the Welch method, as outlined in Appendix A, with a triangular window and 80\% overlap of the segments. Each segment had 100 points with a time discretization of $\Delta t=30$. The total number of segments in the averaging was 90. These choices were seen to be adequate for the current application to accurately resolve the frequencies and wavenumbers of the structures in the flow, exemplified in figure \ref{podcomparisonandoptimalperturbation}. 

The SPOD modes are compared to the flow response to the optimal upstream perturbation, which is calculated by using direct-adjoint power iterations via the boundary layer equations, as in \cite{levin2003exponential}. The objective of such comparison is to determine whether the free-stream turbulence modes are inducing the optimally growing structures, which correspond to streaks for this application. The optimal perturbation is made for a given $(\omega,\beta_k)$, and the comparison made to the most amplified case. The calculation is performed for different streamwise positions and the perturbation which is most amplified with respect to its initial position is chosen for comparison. We have also obtained the flow response to the optimal forcing, adapting the formalism in \cite{levin2003exponential} for resolvent analysis, as shown in Appendix B. The resulting fluctuation at the final integration position is approximately the same for the optimal upstream perturbation and optimal forcing, given that they are both generated at the same streamwise position. 

\begin{figure}
\centering
\includegraphics[width=0.60\textwidth]{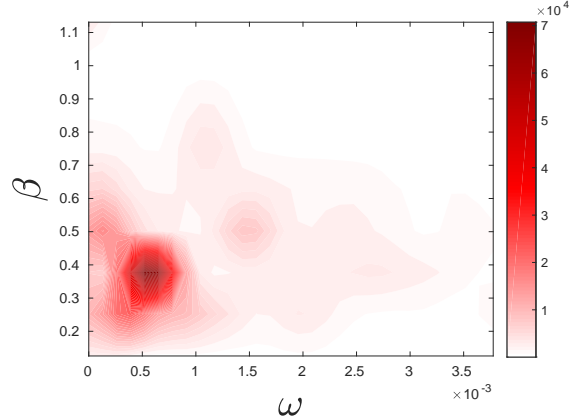}
\caption{Behaviour of the first SPOD eigenvalue as a function of the frequency and spanwise wavenumber, for the Tu 3.0 \% case. Similar characteristics for the most amplified frequency/wavenumber pair are also observed for Tu 3.5\%.}
\label{blablafrequencyspod}
\end{figure}

Figure \ref{podcomparisonandoptimalperturbation} presents the comparison of the leading SPOD mode with the result of the optimal perturbation, which is found to be generated at $x_1 \approx 75$. The behaviour of the first SPOD mode for the streamwise velocity fluctuation in the $(x_2,x_3)$ plane is also shown and highlights the characteristic streaky behaviour of the flow. The calculation was made for $(\omega,\beta_k)=(0,0.37)$, as this corresponds to the most amplified frequency/wavenumber pair, as shown in figure \ref{blablafrequencyspod}.

As highlighted in figure \ref{podcomparisonandoptimalperturbation}, there is a good correspondence between the first SPOD mode of the velocity fluctuations induced by FST and the optimal perturbation. A similar feature had already been observed in other works \citep{luchini2000reynolds}, where the optimal perturbation is seen to be approximately independent of Reynolds number and to match the structures induced by free-stream turbulence modes.

Finally, figure \ref{blablafrequencyspod} presents the behaviour of the first SPOD eigenvalue as a function of the the frequency and transverse wavenumber. Such analysis is necessary for the definition of the $(\omega,\beta_k)$ pair which will be considered in the optimization of the actuator in sections \ref{optimalforcingactuator} and \ref{identifiedactuator}. Although not shown here, the first eigenvalue dominates the dynamics of this flow, being approximately one order of magnitude higher than the subsequent modes. It is clear that the dominating structures are present for $\beta_k \approx 0.37$, which will therefore be targeted by the optimization techniques presented herein.

\section{Actuators}
\label{actuatorstobeconsidered}

A total number of 36 elements is considered in the row of actuation, $u$. Each element adds a body force to the flow with a given spatial support $\mathbf{b}(x_1,x_2,x_3)=(f_{x_1}(x_1,x_2,x_3),$ $f_{x_2}(x_1,x_2,x_3),f_{x_3}(x_1,x_2,x_3))$ which is modulated by a time signal $a_l(t)$,

\begin{equation}
\label{actuationequation}
\mathbf{f}(x_1,x_2,x_3,t)=a_l(t)\mathbf{b}(x_1,x_2,x_3),
\end{equation}

\noindent and the role of the control law is then to determine the time modulation, $a_l(t)$ for each element. Three different actuators will be evaluated which vary in terms of their spatial support.

\subsection{Vertical force only - $f_{x_2}$ actuator}

The first actuator corresponds to a vertical body force only and it seeks to mimic the effect of ring plasma actuators, such as in the works of \cite{kim2016report} and \cite{shahriari2018control} who deal with a plasma actuator with a similar spatial support, acting on the wall-normal direction. The effectiveness of such actuator is related to the lift-up effect, which is a known trigger of streaks \citep{brandt2014lift}. For such actuator, we define the following spatial support, leading to a force only in the wall-normal direction,

\begin{equation}
f_{x_{2}}=\exp\left(-(x_1-x_{1_0})^2/(L_{x_1})^2 -x_2^2/(L_{x_2})^2 -x_3^2/(L_{x_3})^2 \right)
\end{equation}

\noindent with the other components of the forcing equal to zero, $x_{1_0}=325$ corresponding to the position of actuation, $L_{x_1}=3$, $L_{x_2}=5$ and $L_{x_3}=1.5$. The resulting spatial support along the wall-normal direction will be shown in figure \ref{forcesblablabla} in comparison with the two other cases evaluated here. 

\begin{figure}
\begin{center}
\subfigure[]{\includegraphics[width=0.48\textwidth]{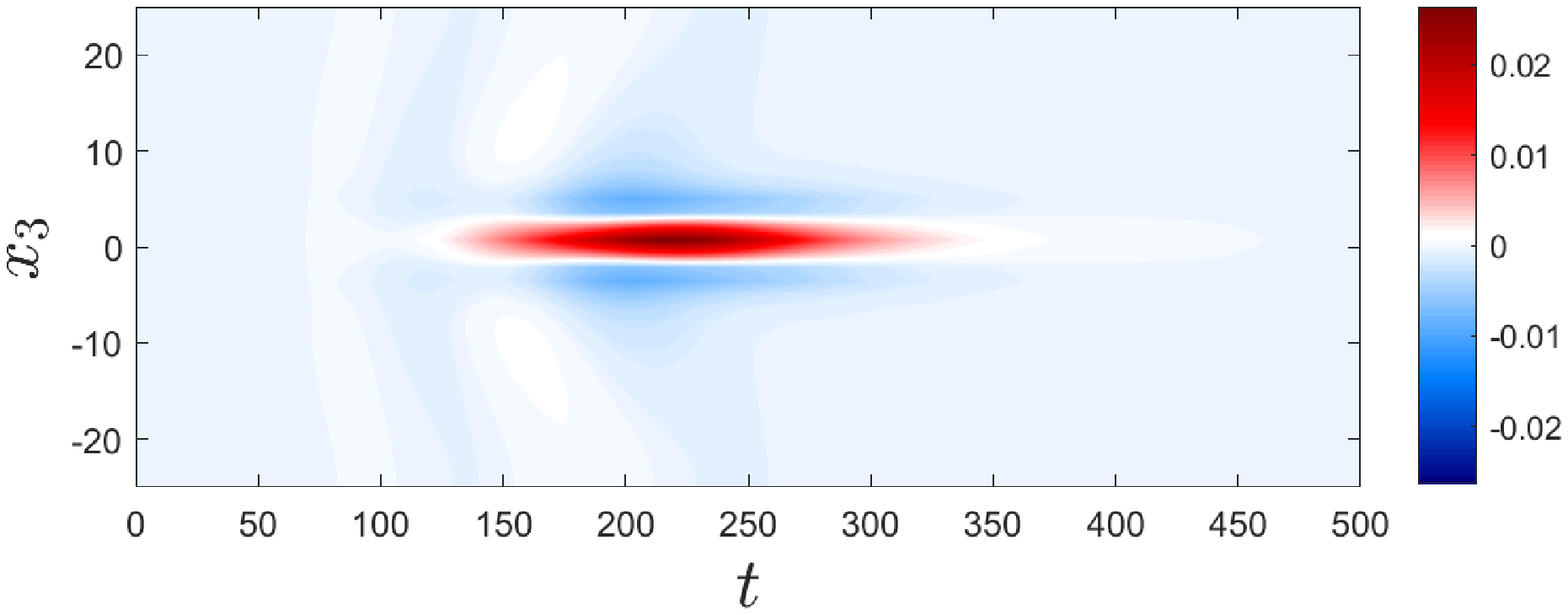}}
\subfigure[]{\includegraphics[width=0.48\textwidth]{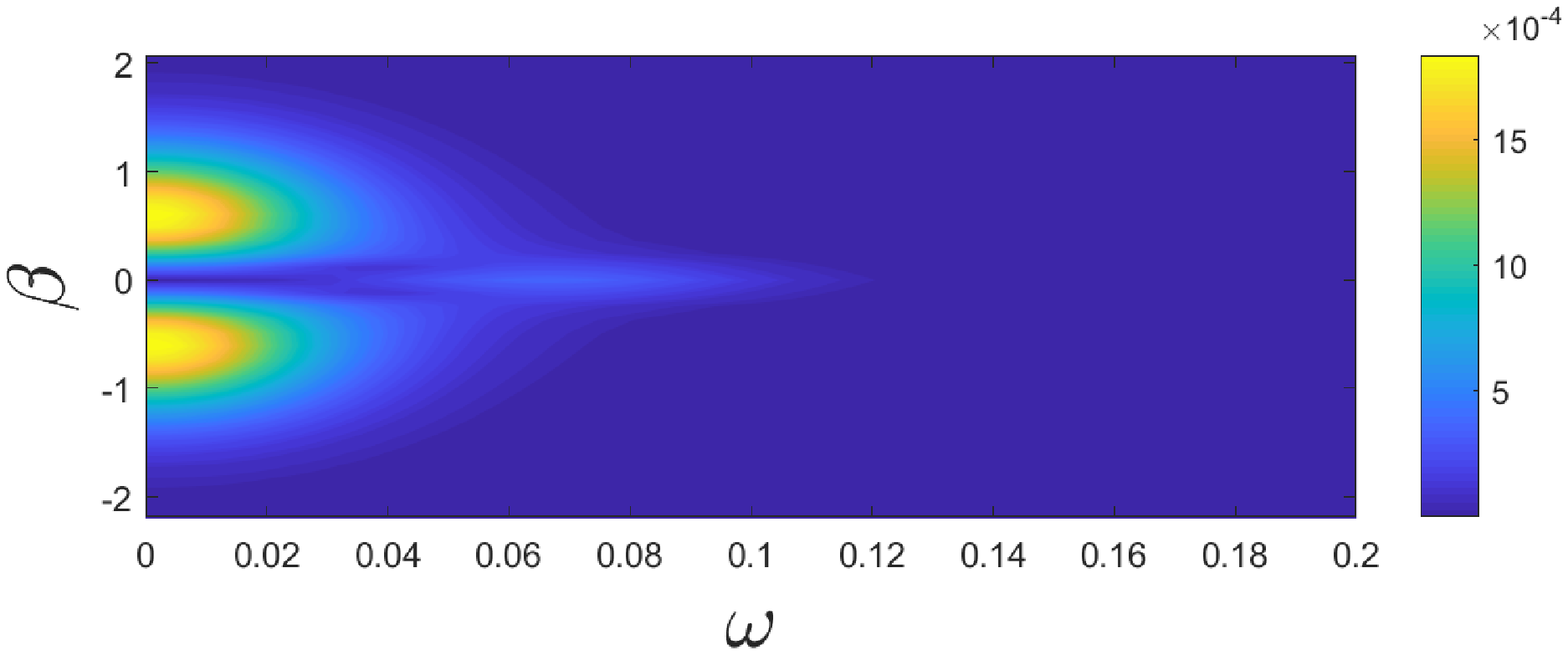}}
\end{center}
   \caption{Impulse response of the blowing actuator, considering wall-shear stress as the measured quantity, in the time (a) and frequency domains (b), respectively.}
  \label{blowingactuatorcharacteristics}
\end{figure}

The impulse response measured at the objective location and its corresponding frequency content are shown in figure \ref{blowingactuatorcharacteristics}. It should be noted that the frequency content of such actuator is concentrated close to $\omega=0$, with a preferable spanwise wavelength $\beta \approx 0.37$, which corresponds to the most amplified streaks generated by the free-stream turbulence. The delay observed in the time-domain is in accordance with the group velocity of such structures and the streamwise separation between the row of actuators and sensors.

\subsection{Optimal forcing actuator}
\label{optimalforcingactuator}

The second actuator to be considered corresponds to a spatial support given in terms of the optimal forcing, which is calculated at the position of actuation, $x_1=325$. It should be noted that such forcing is different from the one considered in section \ref{spodfortheestimation} where the streamwise dependence on the generation of the forcing is also considered. The method to calculate the optimal forcing is outlined in the appendix and corresponds to a modification of the procedure described in \cite{levin2003exponential}, using adjoint methods for constrained optimization; the goal is to obtain the forcing that leads to the highest energy gain at the position of objective, at $x_1=400$, for the most amplified spanwise wavenumber, $\beta \approx 0.37$.

The actuation is restricted to spatially localized upstream areas by inclusion of a Gaussian mask in the optimization procedure. This avoids a spatially extended forcing which would be impractical in experimental applications, for example. As shown in Appendix B, the spanwise spatial support is imposed by a Gaussian function given by $\exp(-x_3^2/(L_{x_3})^2)$, with $L_{x_3}=1.5$. The resulting spatial support along the wall-normal direction is shown in figure \ref{forcesblablabla} for the span and wall-normal components; the contribution of the streamwise forcing is irrelevant, as it is comparably inefficient for the generation of streaks.

\begin{figure}
\begin{center}
\subfigure[]{\includegraphics[width=0.48\textwidth]{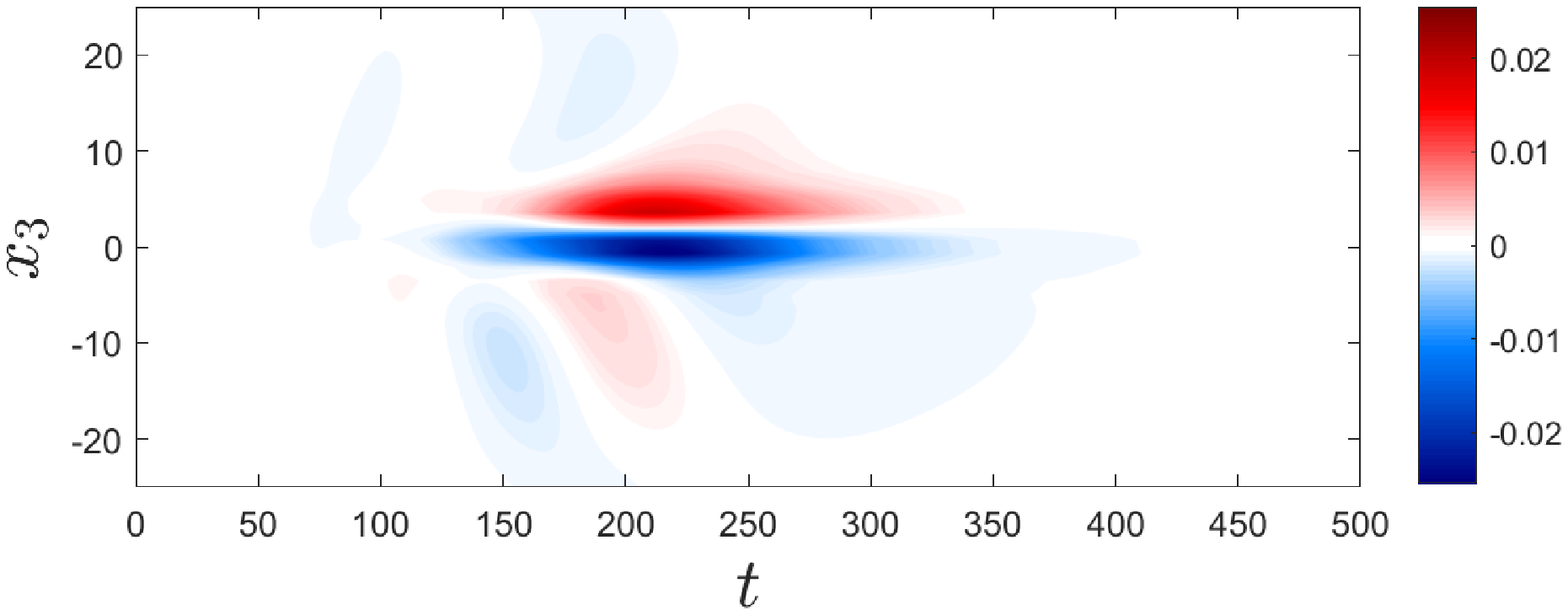}} 
\subfigure[]{\includegraphics[width=0.48\textwidth]{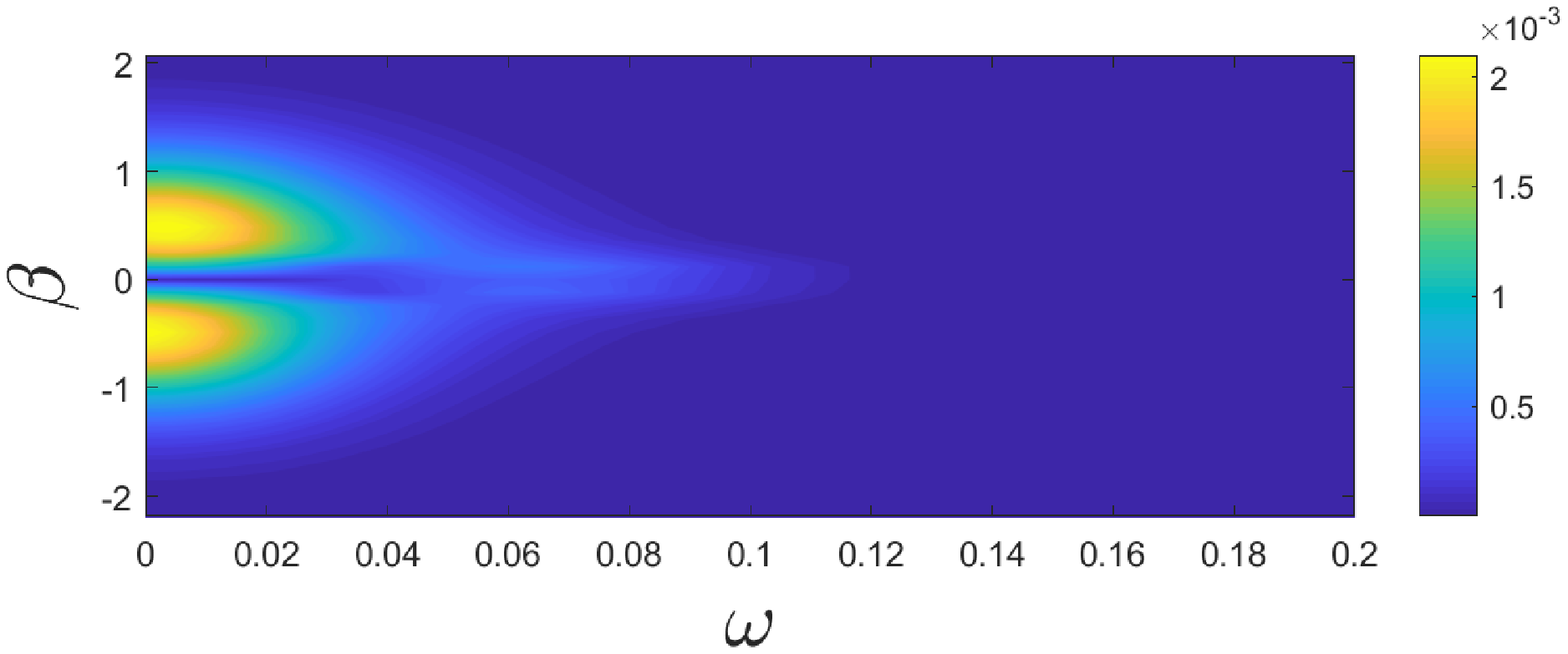}}
\end{center}
   \caption{Time (a) and frequency-domain (b) behaviour of the impulse response of the optimal forcing actuator considering wall-shear stress as the measured output quantity.}
  \label{freqandtimedataofoptimalguy}
\end{figure}

The corresponding impulse response in the time and frequency domains are shown in figure \ref{freqandtimedataofoptimalguy}. As before, the most amplified streaks are located at $\beta \approx 0.37$, in accordance with the performed optimization.

\subsection{Identified actuator}
\label{identifiedactuator}

\begin{figure}
\begin{center}
\subfigure[]{\includegraphics[width=0.48\textwidth]{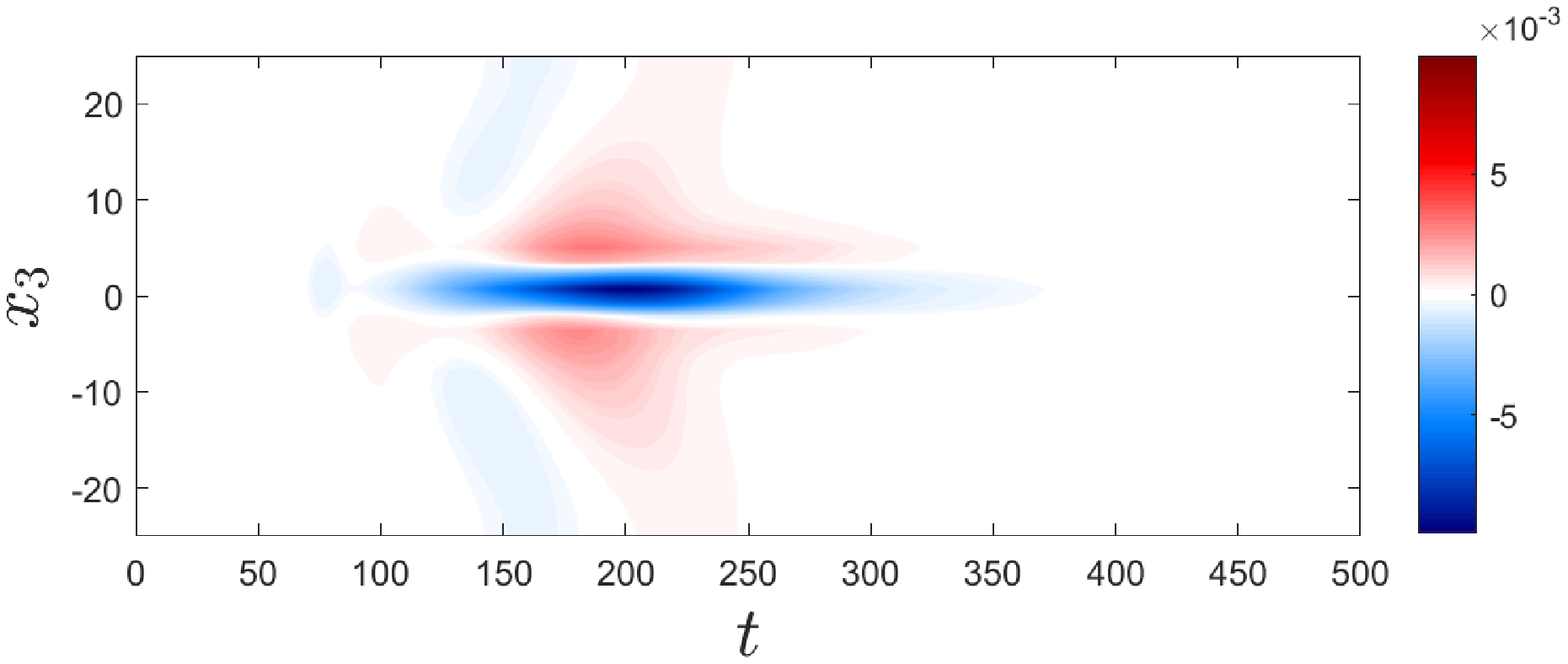}} 
\subfigure[]{\includegraphics[width=0.48\textwidth]{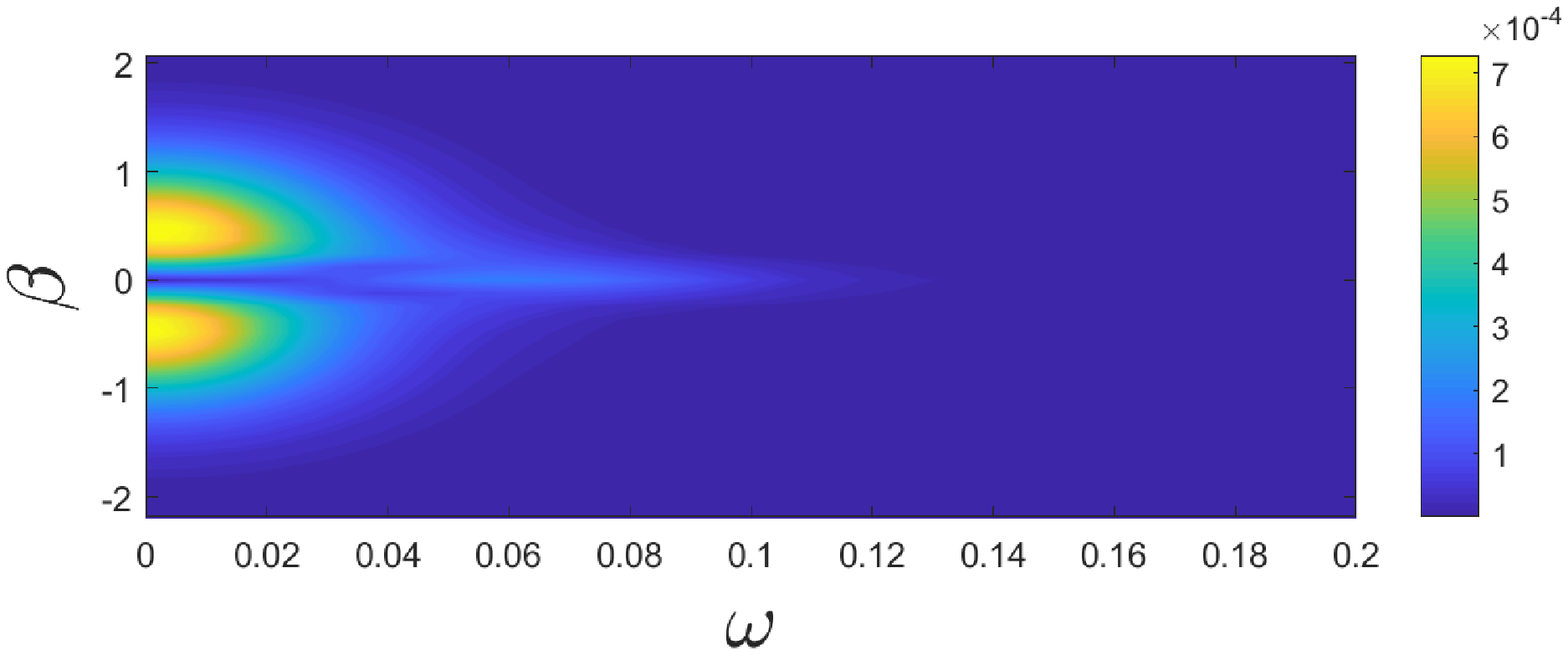}}
\end{center}
   \caption{Time (a) and frequency-domain (b) behaviour of the impulse response of the identified actuator, wall-shear stress as considered as the measured output quantity.}
  \label{Impulse_freq_Lx6_y02_Ly6_beta0377}
\end{figure}

Finally, the third actuator to be considered is calculated targeting the specific shape of the structures present at the objective position $x_1=400$, given in terms of their first SPOD mode for the most amplified $(\omega,\beta_k)$ pair, as described in section \ref{spodfortheestimation}. The actuator will be referred to as ``identified'' as it targets the structures which were previously identified at the position of objective. This procedure is also outlined in Appendix B and it is inspired in the work of \cite{tissot2017sensitivity}. This actuator is expected to be the most efficient, as it targets the specific structures present in the flow and should therefore lead to their best cancellation, in accordance with the physical mechanisms behind active flow control. The resulting impulse response is shown in figure \ref{Impulse_freq_Lx6_y02_Ly6_beta0377}. The optimization was performed exclusively with the wall-normal and spanwise direction forcings, which should dominate the generation of streaks.

As before, the maximum of the frequency content is consistent with the targeted streaks; the consideration of the spanwise component of the forcing causes the $(x_3,t)$ behaviour to be non-symmetric along the $x_3$ direction.

\subsection{Comparison of the different forcings}

\begin{figure}
\begin{center}
\subfigure[]{\includegraphics[width=0.45\textwidth]{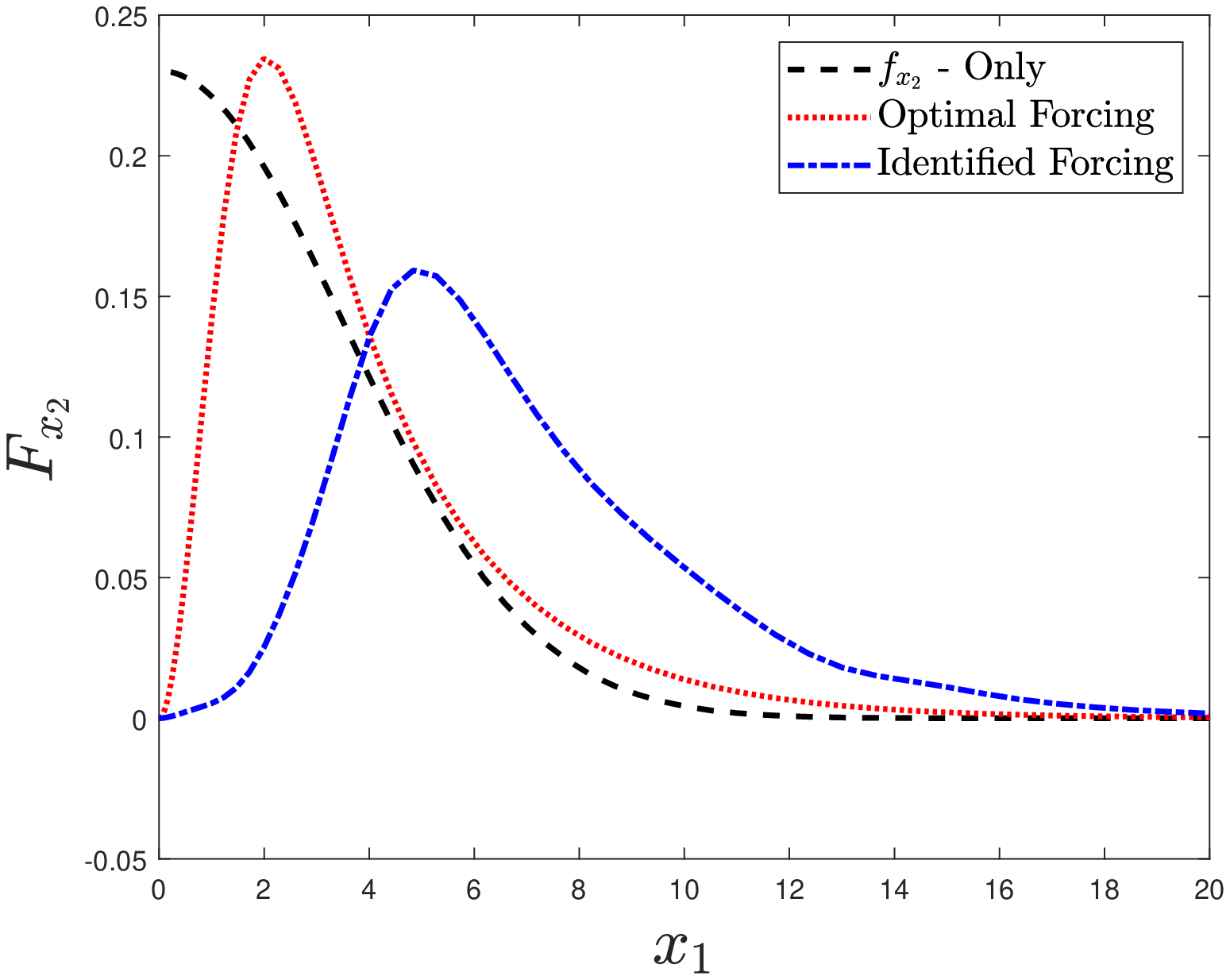}} 
\subfigure[]{\includegraphics[width=0.45\textwidth]{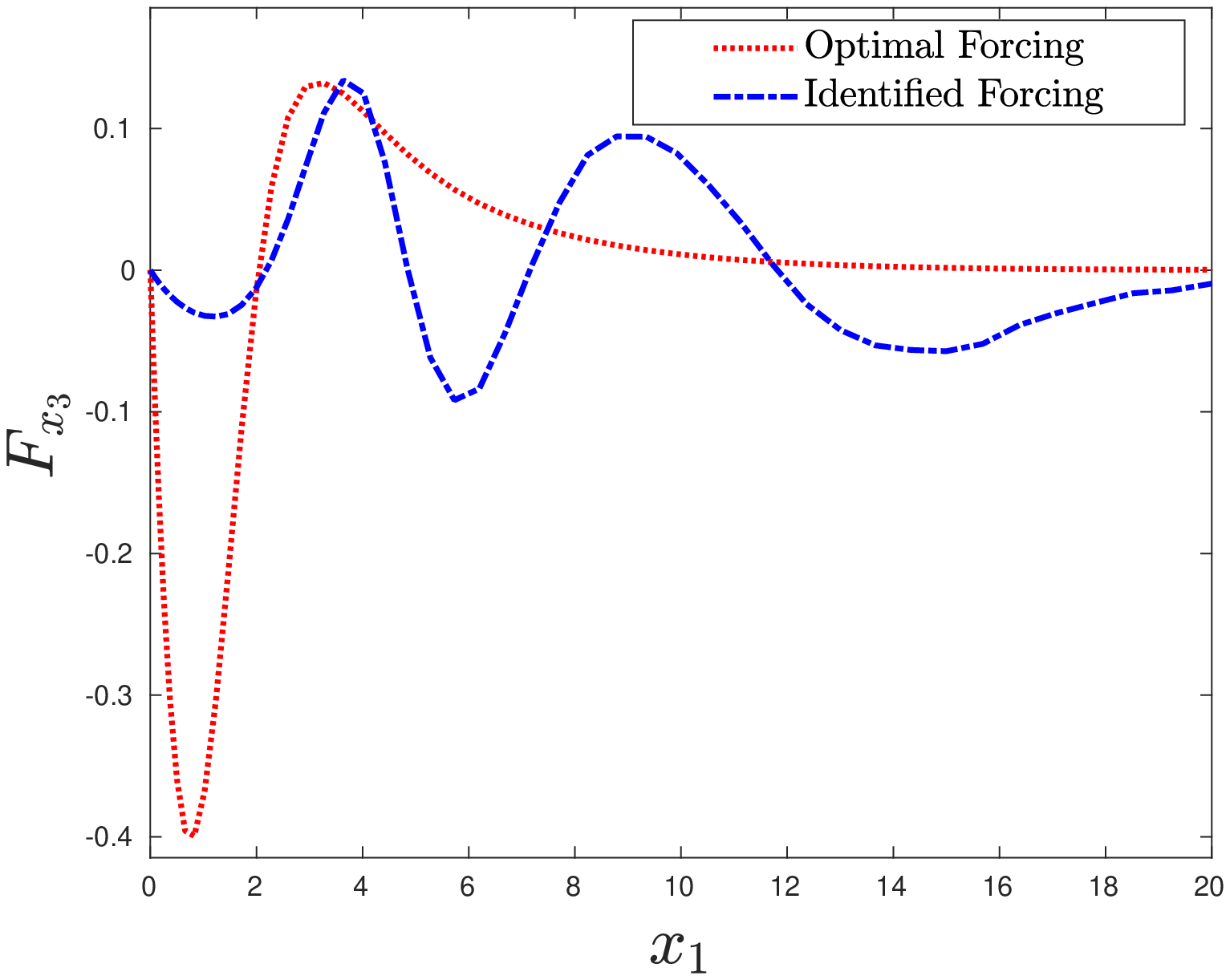}}
\end{center}
   \caption{Spatial support of the three forcings considered along the wall-normal direction for the streamwise (a) and wall-normal (b) components, respectively.}
  \label{forcesblablabla}
\end{figure}

The main difference on the spatial support of the forcings is on their wall-normal behaviour, as the same Gaussian mask was considered in the span and streamwise directions. The three different cases are shown in figure \ref{forcesblablabla} for the wall-normal and spanwise components. The spatial support was normalized such that the energy content of the different forcings is the same.

The two optimization techniques lead to the typical behaviour for the optimal forcing shape as in \cite{monokrousos2010global}. The main difference between the optimal forcing and SPOD-based optimization is that the latter presents a peak at higher wall-normal positions, a feature which is seen to be related to the streaks the actuator generates.

\begin{figure}
\begin{center}
\includegraphics[width=0.75\textwidth]{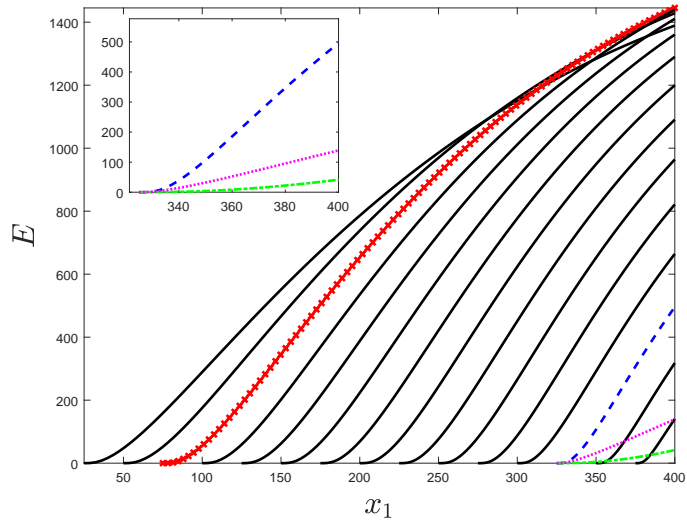}
\end{center}
   \caption{Comparison of the energy fluctuation as a function of the streamwise position for the different forcings considered; optimal forcing calculated for different streamwise positions (black solid), optimal forcing at the most amplified position (red crosses), optimal forcing calculated at the position of actuation (blue dashed), vertical forcing (pink dotted) and SPOD-based identification (green dash-dotted). A zoom of the area of interest ($x_1 \geq 325$) is shown in the inset.}
  \label{optimalforcingschemedifferentcases}
\end{figure}

Finally, the energy $E$ of the fluctuation, 

\begin{equation}
\label{energyequationalongthestreamwisedirection}
E(x_1,\omega,\beta_k)=\int_0^{\infty}|\hat{\mathbf{q}}(x_1,x_2,\omega,\beta_k)|^2\mathrm{d}x_2
\end{equation}

\noindent resulting from the application of these forcings is shown in figure \ref{optimalforcingschemedifferentcases}. The result is compared to the calculation of the optimal forcing at different streamwise locations. There is a strong dependence of the fluctuation energy on the position where the optimization is performed, as previously observed in other works \citep{levin2003exponential}. The optimal position for the generation of the optimal forcing is at  $x_1 \approx 75$ and the fluctuation resulting from such forcing approximately matches the FST induced streak at the objective position for control ($x_1 \approx 400$), as previously seen in figure \ref{podcomparisonandoptimalperturbation}.

The optimal forcing calculated at the $x_1=325$, where the actuation is actually performed for control, leads to a much higher energy at the objective position when compared to the other two approaches. However, as it will be shown later, it leads to a thinner streak when compared to the actual structures inside the boundary layer.

\section{Results}
\label{resultsfortransitiondelaymainly}

The attention will be focused on two turbulence intensity cases, $Tu=3.0\%$ and $3.5\%$, both of which present challenging scenarios on which transition to turbulence will be induced by the free-stream disturbances and where some nonlinearity is already present in the sensor/actuator region, posing a limitation in the accuracy of the considered reduced-order models. The same kernels and actuators, designed for $Tu=3.0\%$ were used for both cases which also allows an evaluation of the robustness of the controllers and optimization methods considered. Different kernels were calculated for each actuator, a necessary step since the actuators present significant differences in their impulse responses, illustrated in figures \ref{blowingactuatorcharacteristics} to \ref{Impulse_freq_Lx6_y02_Ly6_beta0377}. 

\begin{figure}
\begin{center}
\subfigure[Performance index for $Tu=3.0\%$]{\includegraphics[width=0.85\textwidth]{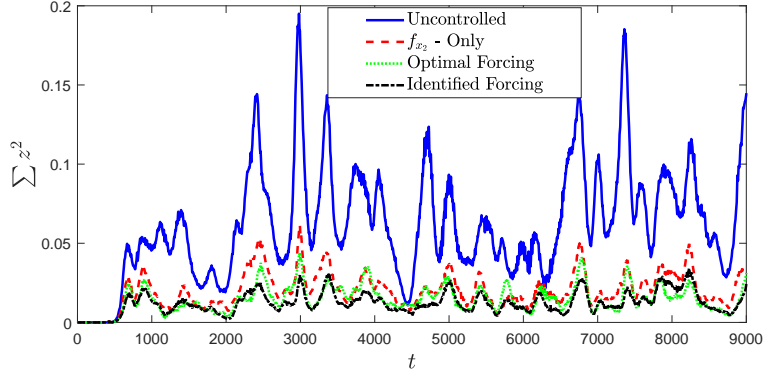}} 
\subfigure[Performance index for $Tu=3.5\%$]{\includegraphics[width=0.85\textwidth]{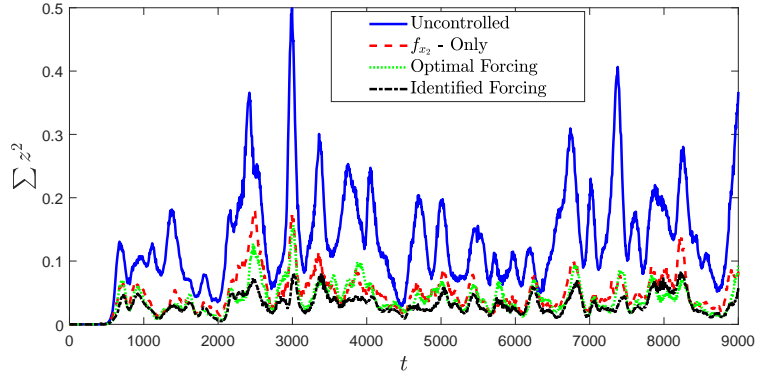}} \\
\end{center}
   \caption{Performance indices for the two turbulence intensity cases evaluated.}
  \label{figurewithperformancemetrics}
\end{figure}

Figure \ref{figurewithperformancemetrics} presents the performance indices, which consider the mean square values of the output, for the two cases evaluated considering the three different actuators. It should be noted that the performance index takes into account exclusively the wall-shear stress and therefore does not represent a metric for the disturbances throughout the boundary layer. On the other hand this parameter serves as a good evaluation of the effectiveness of the controllers themselves, as their single objective is the minimization of the wall-shear fluctuations at the objective position.

All three methods present an adequate attenuation of the objective, with the identified actuator outperforming the other two. Table \ref{tableofcontroldatafsadfsafadf} summarizes the average reduction for the different cases. The better performance observed for the $Tu=3.0\%$ case is related to a better accuracy of the reduced-order models, where less non-linear effects are present in comparison to $Tu=3.5\%$.

\begin{table}
\centering
\begin{tabular}{ccc} \toprule
Actuator                       & $Tu=3.0\%$       & $Tu=3.5\%$ \\ \midrule
$f_{x_2}$ - Only                & $z_{con}^2/z_{unc}^2=0.33$           & $z_{con}^2/z_{unc}^2=0.34$   \\ 
Identified  		             & $z_{con}^2/z_{unc}^2=0.21$	   & $z_{con}^2/z_{unc}^2=0.25$  \\ 
Optimal Forcing  	 & $z_{con}^2/z_{unc}^2=0.25$ 	   & $z_{con}^2/z_{unc}^2=0.33$   \\ \bottomrule
\end{tabular}
\caption{Summary of the closed-loop cases evaluated.}
\label{tableofcontroldatafsadfsafadf}
\end{table}

In order to evaluate the energy spent in actuation, the following metric, which is related to the energy budget for actuation, is defined;

\begin{equation}
\label{energymetric}
E_u(t)=\int_{x_{1_0}}^{x_{1_{max}}}\int_{x_{2_0}}^{x_{2_{max}}}\int_{x_{3_0}}^{x_{3_{max}}} |\mathbf{b}(x_1,x_2,x_3)a(t)|^2 \mathrm{d}x_1\mathrm{d}x_2\mathrm{d}x_3.
\end{equation}

\begin{figure}
\begin{center}
\subfigure[Energy budget for $Tu=3.0\%$]{\includegraphics[width=0.85\textwidth]{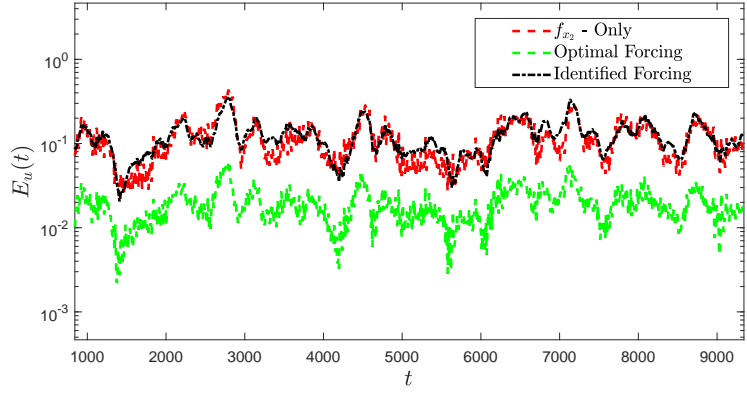}} 
\subfigure[Energy budget for $Tu=3.5\%$]{\includegraphics[width=0.85\textwidth]{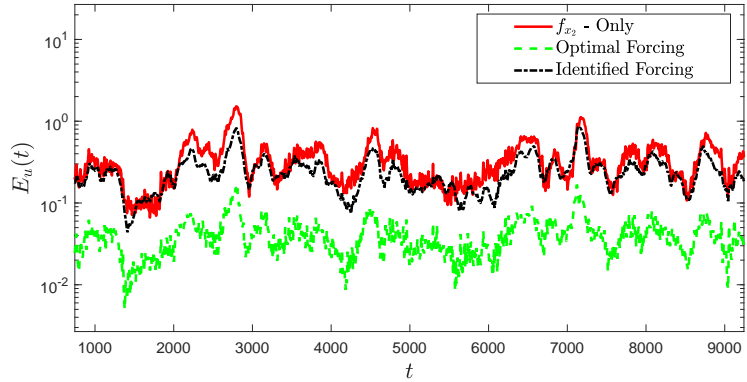}} \\
\end{center}
   \caption{Energy budget for the different turbulence intensity and actuators evaluated.}
  \label{energybudgetdsafsd}
\end{figure}

\noindent This metric considers both the amplitude modulation $a(t)$, which is calculated by the control law, and the spatial support of the forcing $\mathbf{b}$. The behaviour of $E_u(t)$ for the different cases is shown in figure \ref{energybudgetdsafsd}. The energy budget of the optimal forcing actuator is about one order of magnitude lower than the one corresponding to the other two cases, a fact which is related to the streaks induced by it presenting the highest possible growth rates for the specific position where they are generated, as illustrated in figure \ref{optimalforcingschemedifferentcases}.

\begin{figure}
\begin{center}
\subfigure[Friction coefficient, $Tu=3.0\%$]{\includegraphics[width=0.95\textwidth]{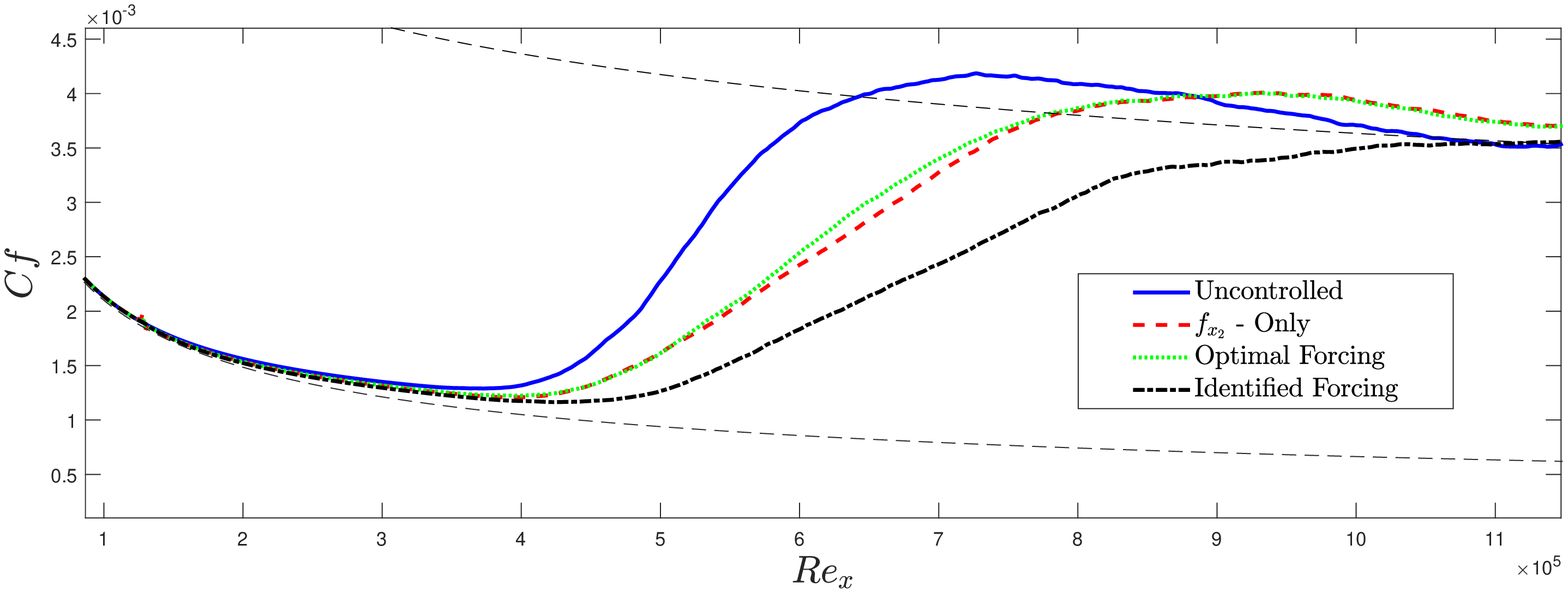}} 
\subfigure[Maximum RMS values along the wall-normal, $Tu=3.0\%$]{\includegraphics[width=0.95\textwidth]{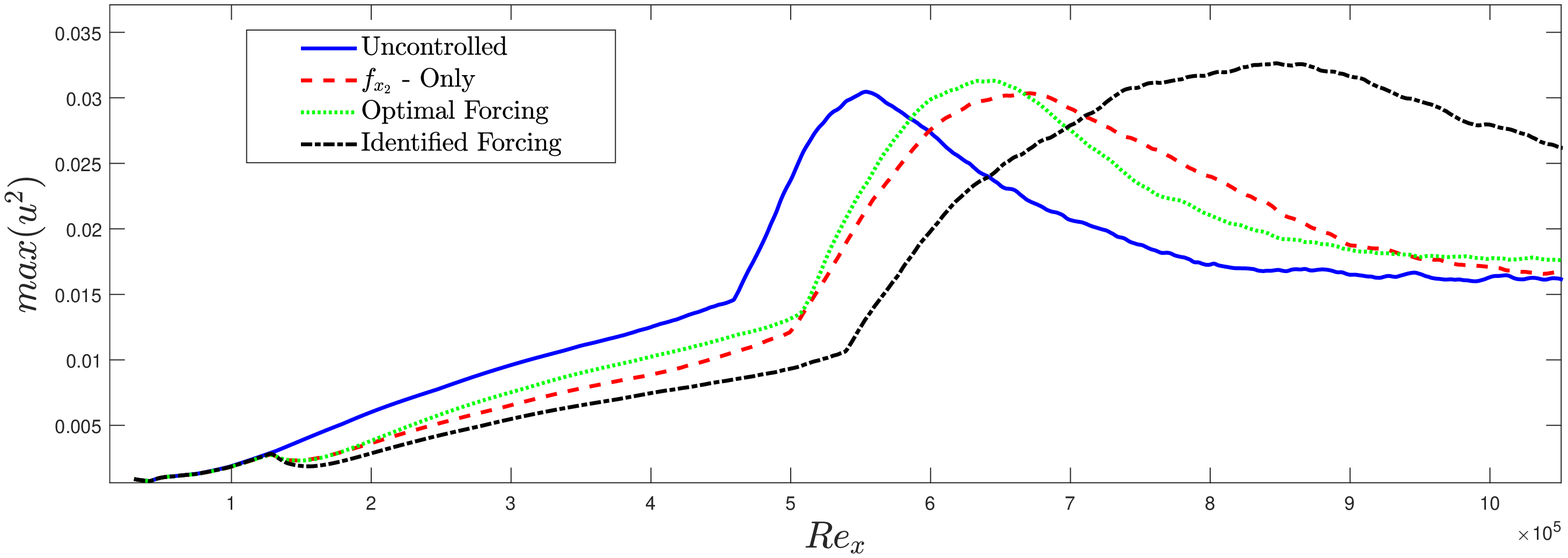}} \\
\subfigure[Friction coefficient, $Tu=3.5\%$]{\includegraphics[width=0.95\textwidth]{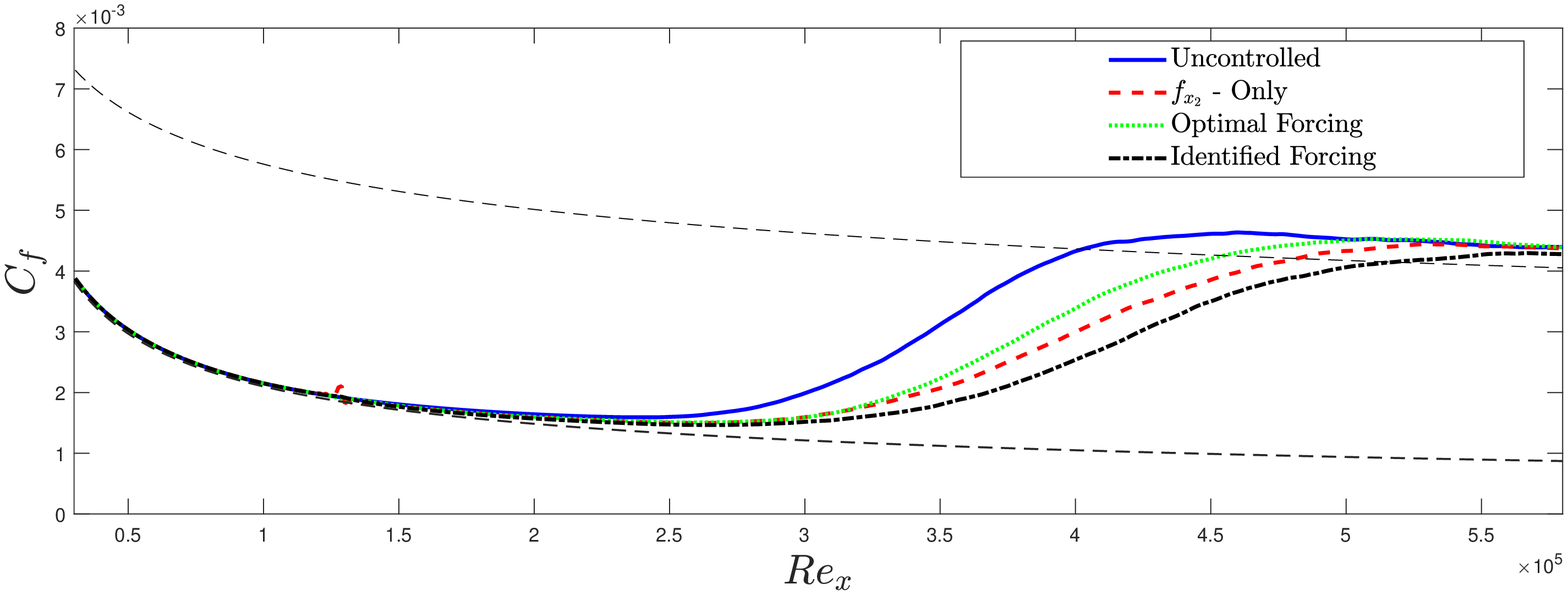}} 
\subfigure[Maximum RMS values along the wall-normal direction, $Tu=3.5\%$]{\includegraphics[width=0.95\textwidth]{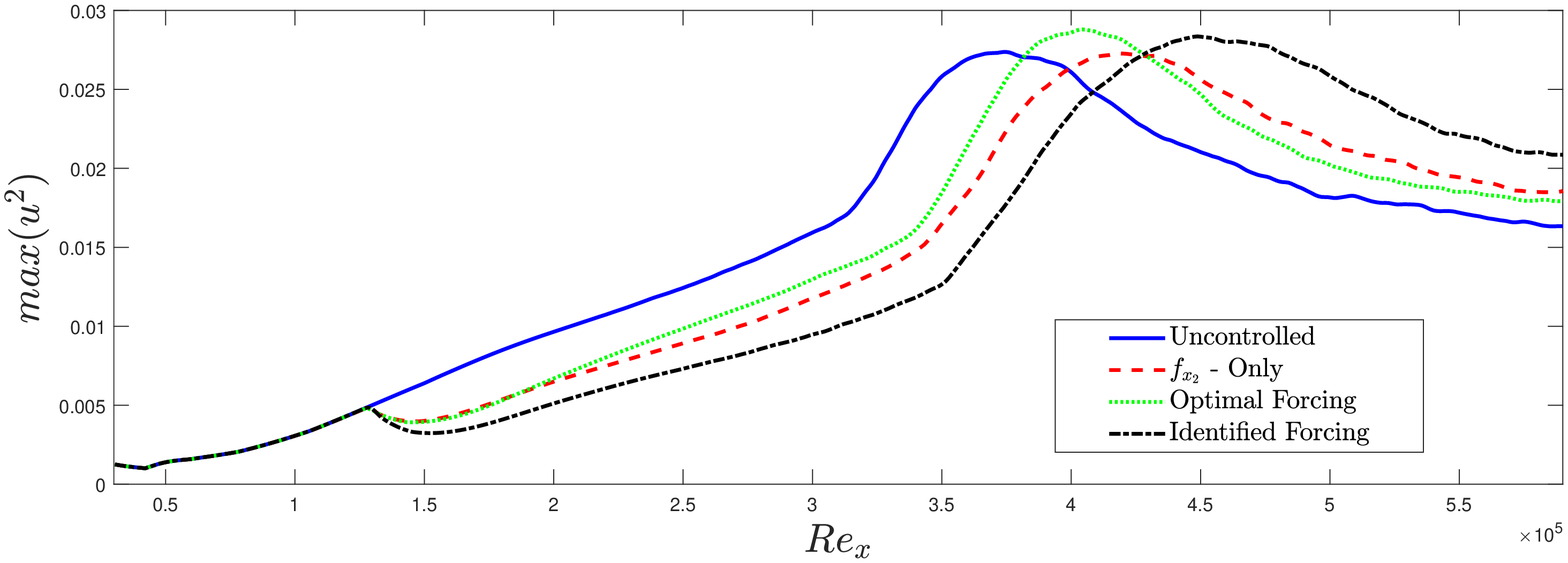}}
\end{center}
   \caption{Friction coefficient and maximum rms values for the streamwise velocity fluctuation and the different actuation schemes considered in this work. The dashed line in the friction coefficient plots gives reference values for the laminar and turbulent cases respectively.}
  \label{figurewiththeimportantresults}
\end{figure}

\begin{figure}
\begin{center}
\subfigure[$Tu=3.0\%$, uncontrolled case]{\includegraphics[width=0.49\textwidth]{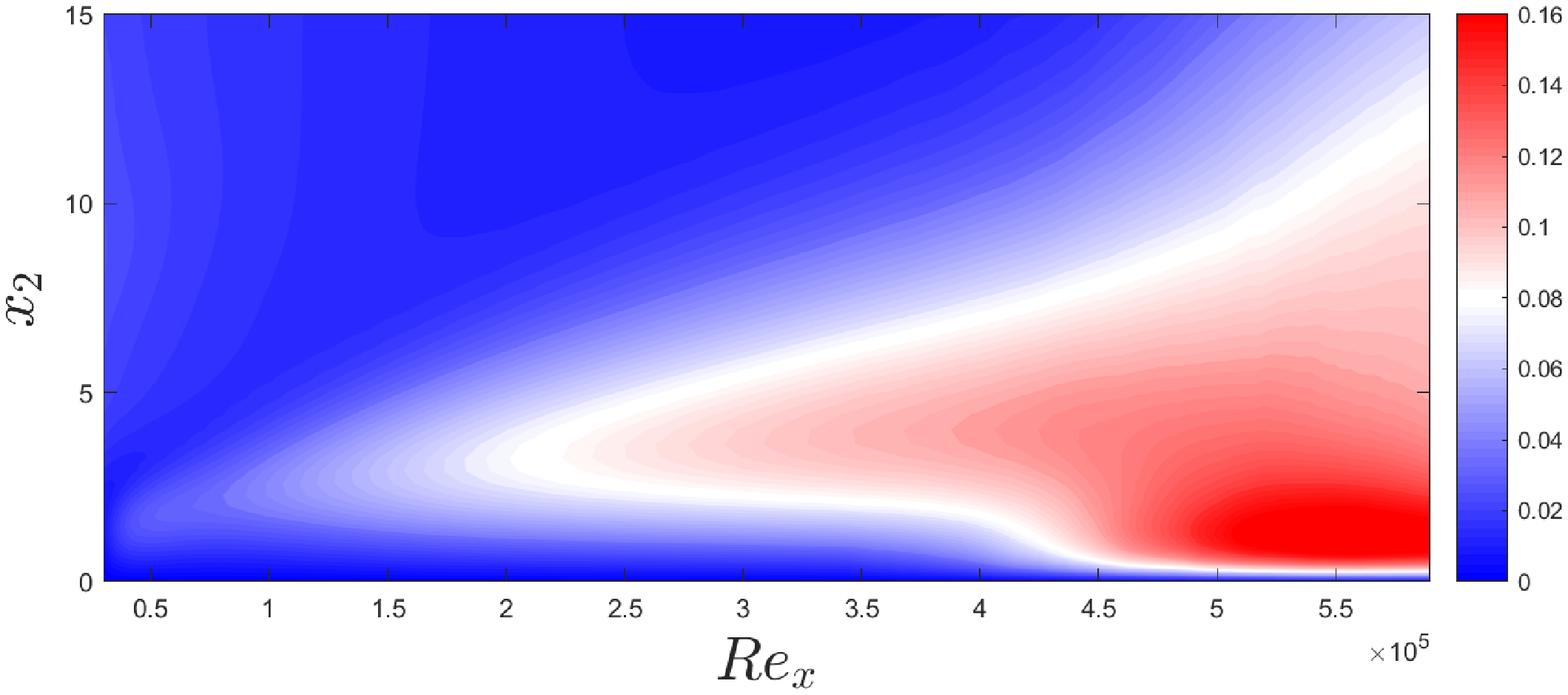}} 
\subfigure[$Tu=3.0\%$, controlled with $f_{x_2}$ forcing]{\includegraphics[width=0.49\textwidth]{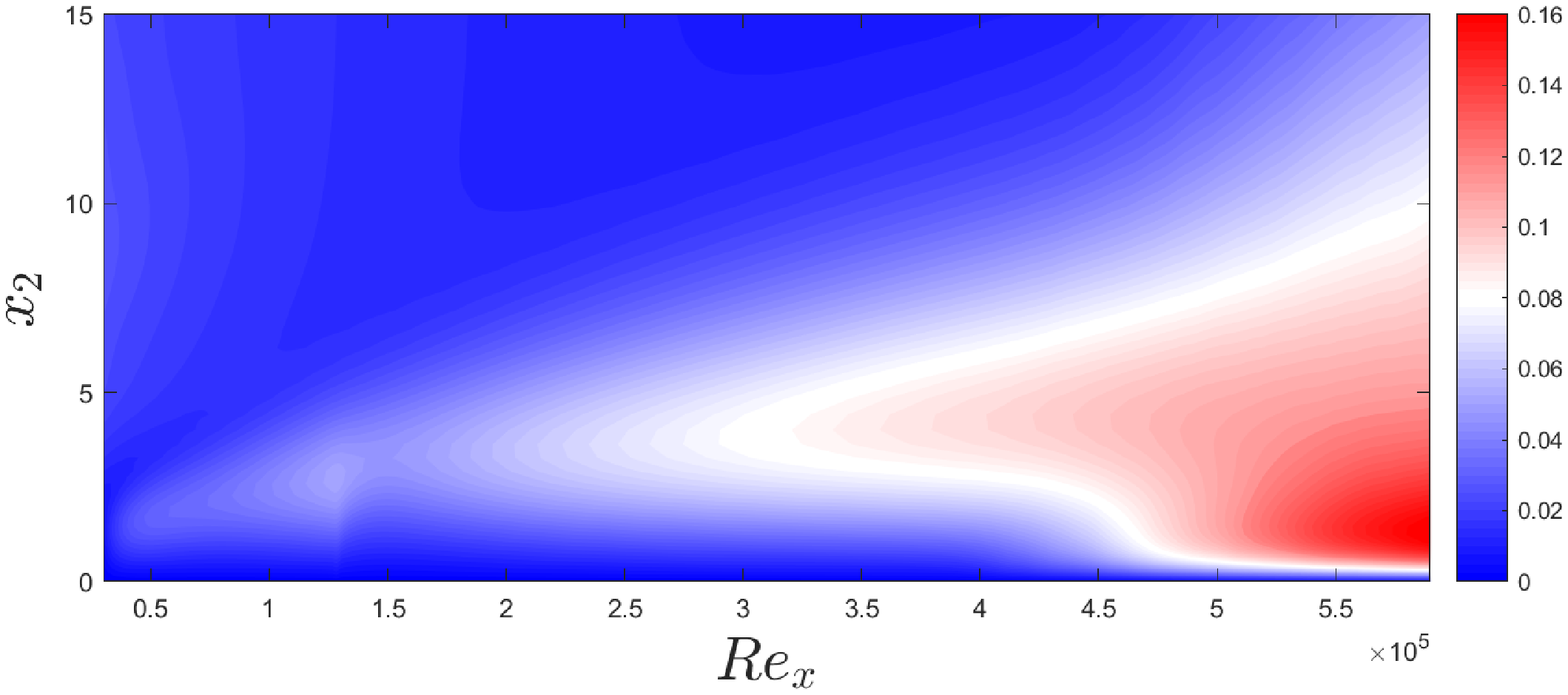}} \\
\subfigure[$Tu=3.0\%$, controlled with the optimal forcing actuator]{\includegraphics[width=0.49\textwidth]{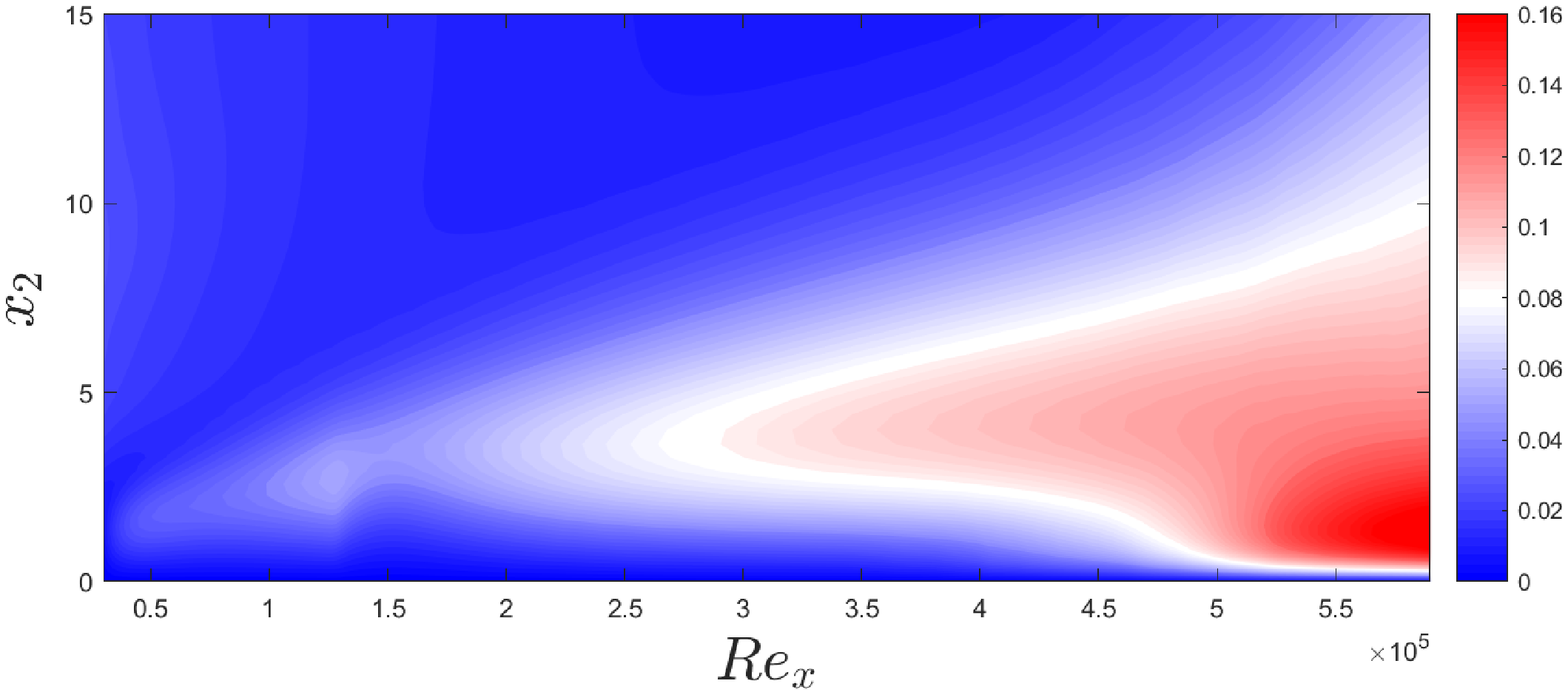}} 
\subfigure[$Tu=3.0\%$, controlled with the identified actuator]{\includegraphics[width=0.49\textwidth]{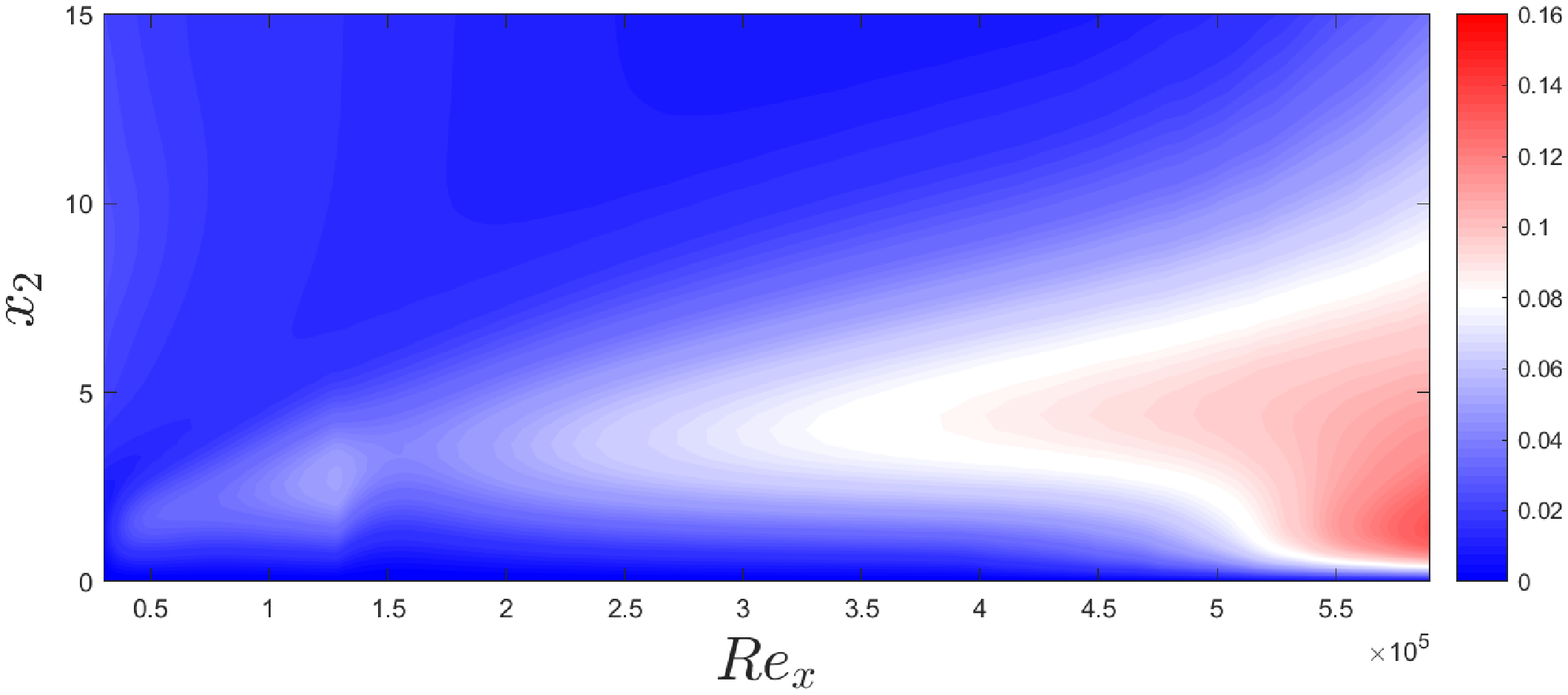}}
\end{center}
   \caption{Behaviour of the RMS of the streamwise velocity fluctuation for the uncontrolled and different controlled scenarios evaluated.}
  \label{figurewiththermsvalues}
\end{figure}

\begin{figure}
\begin{center}
\subfigure[$Re_x=150000$]{\includegraphics[width=0.45\textwidth]{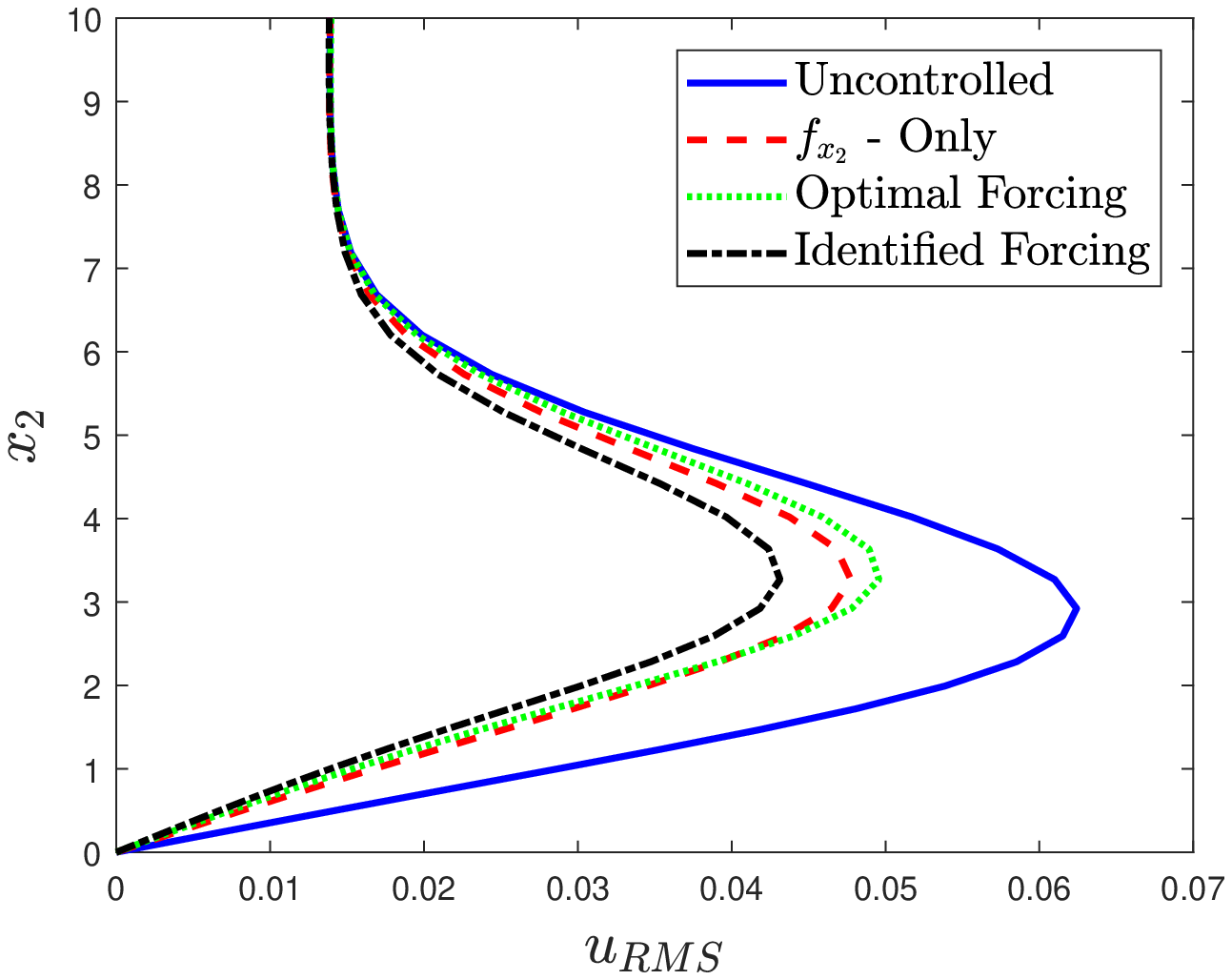}} 
\subfigure[$Re_x=200000$]{\includegraphics[width=0.45\textwidth]{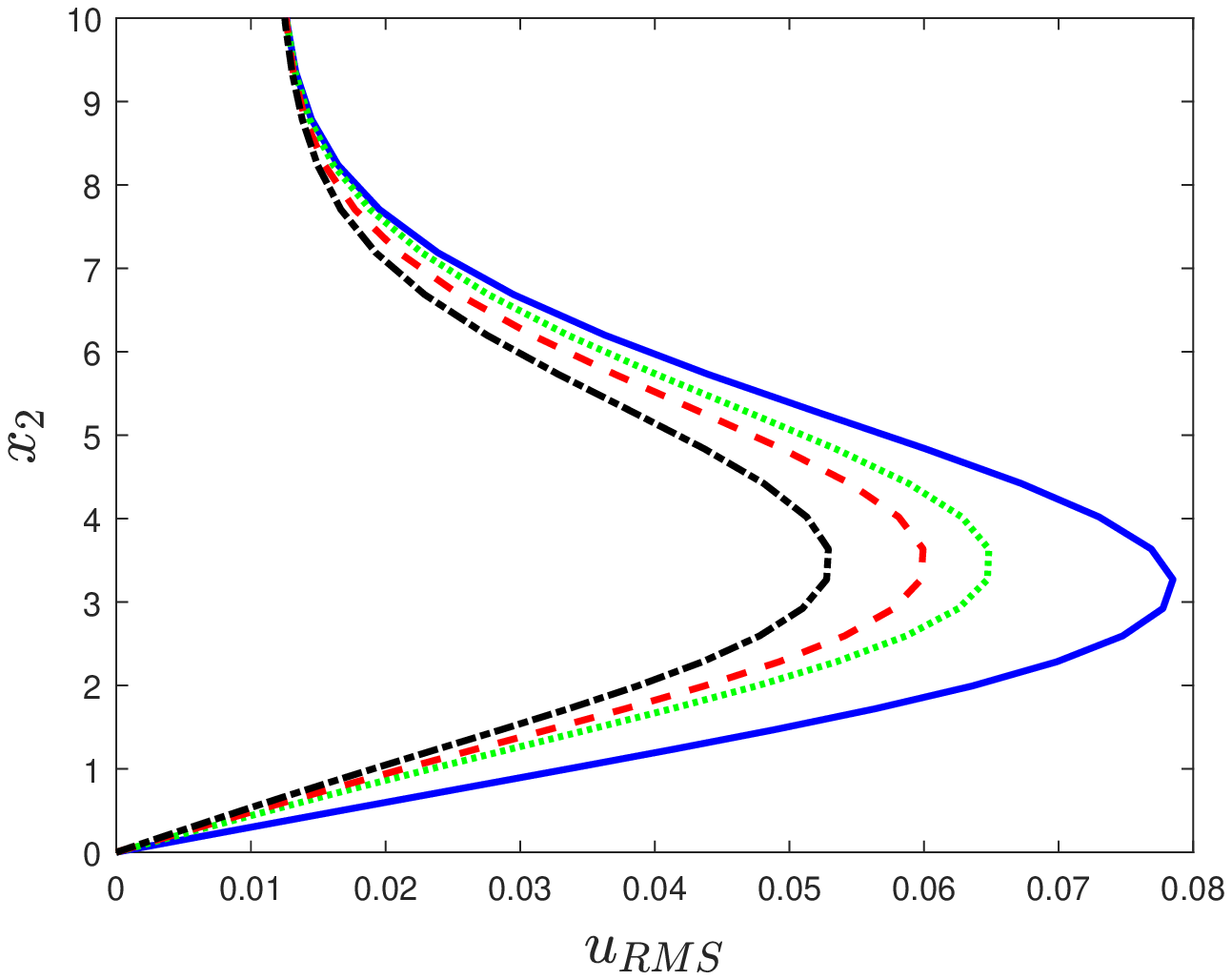}} \\
\subfigure[$Re_x=250000$]{\includegraphics[width=0.45\textwidth]{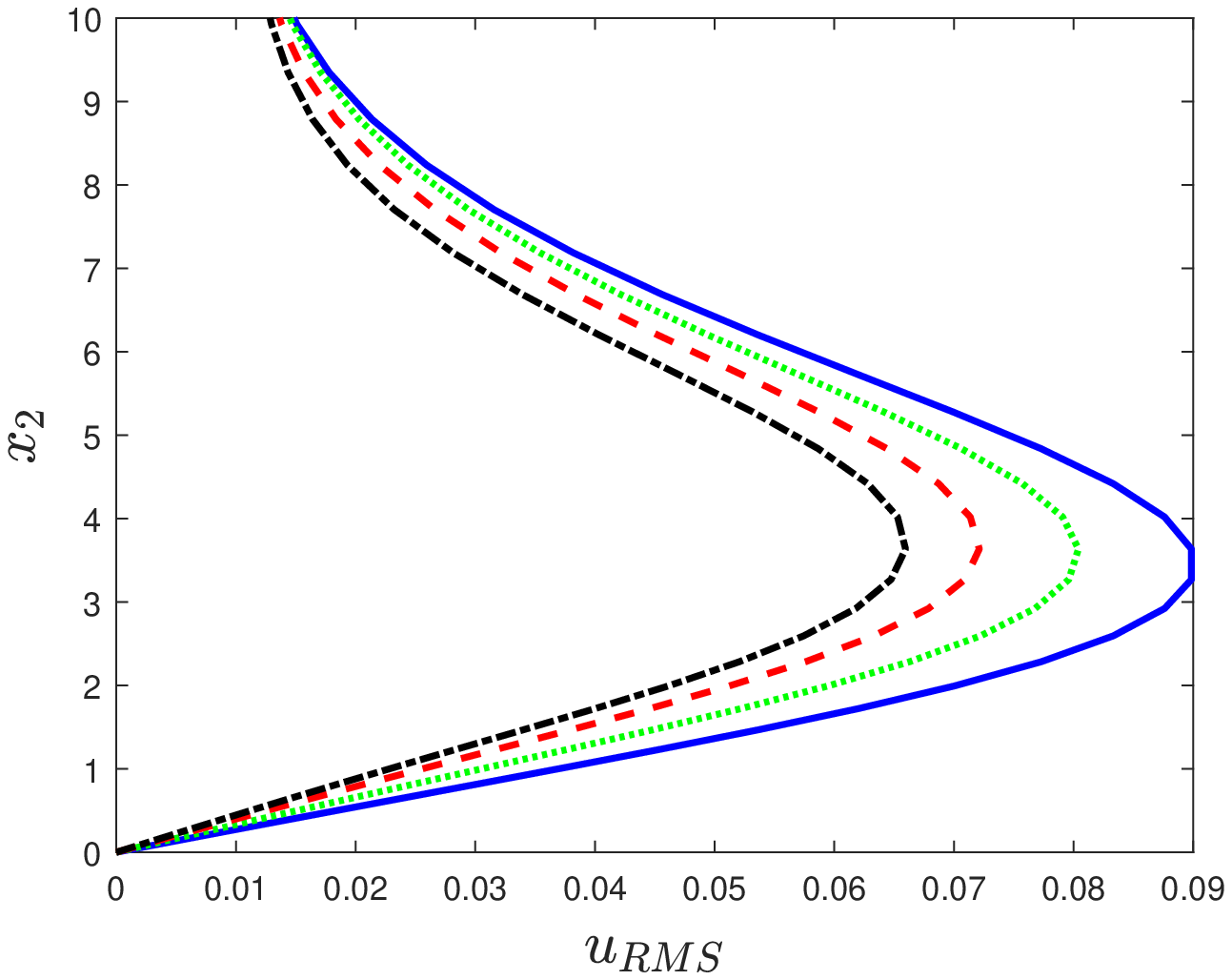}} 
\subfigure[$Re_x=300000$]{\includegraphics[width=0.45\textwidth]{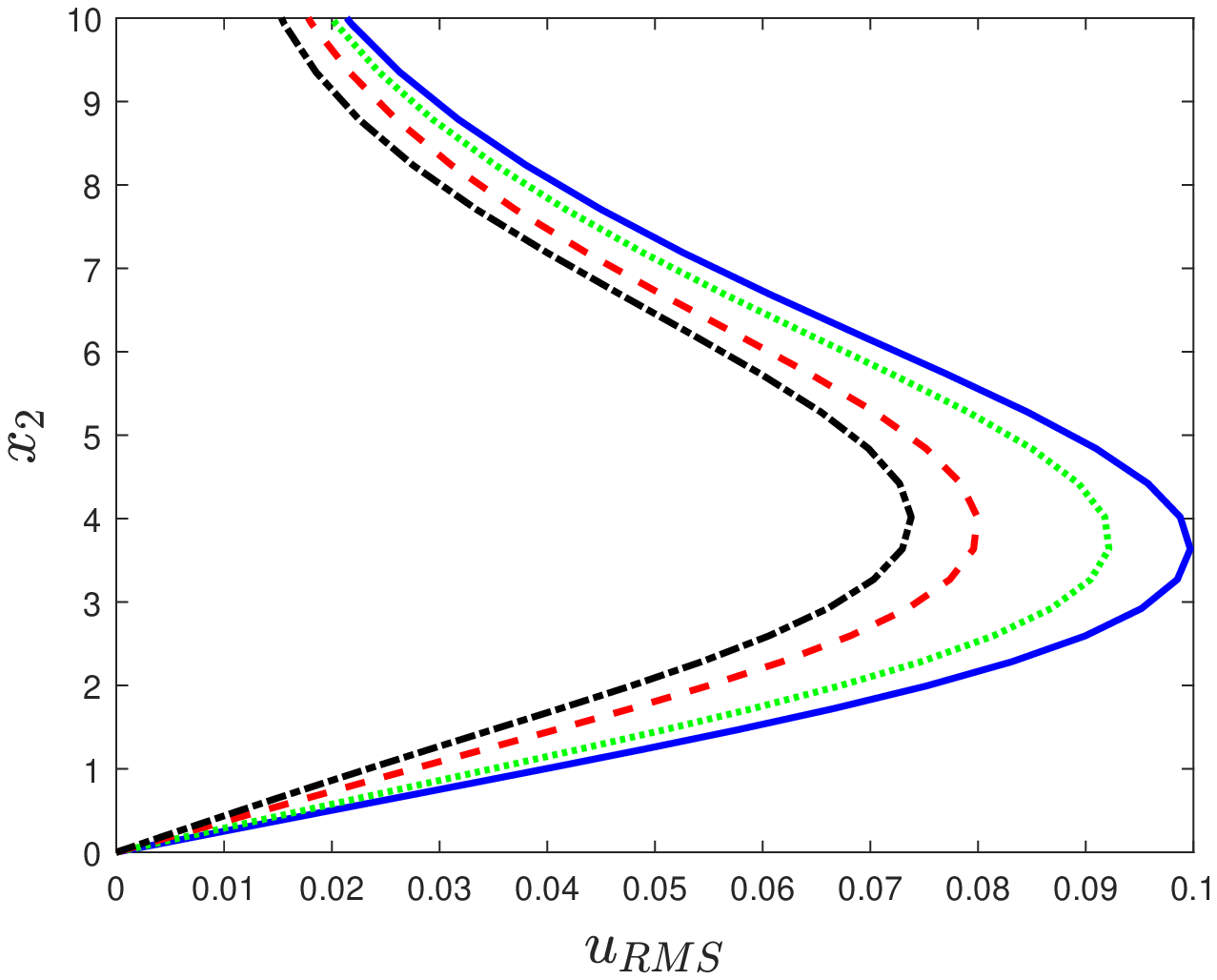}}
\end{center}
   \caption{RMS value of the streamwise velocity fluctuations at four streamwise positions as a function of the wall-normal direction for the uncontrolled and different actuators evaluated in this work. $Re_x=150000$ corresponds to the objective position.}
  \label{figurewiththermsvaluesatdifferentstreamwisepositions}
\end{figure}

We now evaluate the effectiveness of closed-loop control with the different actuators in delaying transition. Figure \ref{figurewiththeimportantresults} shows the friction coefficient $C_f$ and maximum root-mean square (RMS) values for the three evaluated actuators for the turbulence intensity levels of 3.0\% and 3.5 \%. The corresponding behaviour of the RMS values of the streamwise velocity fluctuation in the $(x_1,x_2)$ plane are shown in figure \ref{figurewiththermsvalues}, for the $Tu=3.0\%$ case and in figure \ref{figurewiththermsvaluesatdifferentstreamwisepositions} four streamwise positions are shown in order to better highlight the effect of the different actuation schemes; similar results are also observed for the higher turbulence intensity value. 

The identified actuator considerably outperforms the other two on what concerns the delay in transition, a feature observed from the friction coefficient and maximum RMS values in figure \ref{figurewiththeimportantresults}, which take much longer to increase to values typical of turbulent boundary layers. The RMS values in figure \ref{figurewiththermsvalues} indicate that the effect of the identified actuator is more extended along the wall-normal direction, since, at the position of actuation, there is a significant decrease of RMS levels throughout the boundary layer. Furthermore, the actuation energy of the identified actuator is similar to that of the $f_{x_2}$-only actuator and it is also more robust to the evaluated changes in turbulence intensity, leading also to significant delays in transition for the two evaluated cases.

As for the optimal forcing actuator, although it is capable of delaying the transition in the two FST intensity cases, it presents the lowest performance in terms of the transition delay. The RMS values indicate that its effect is more localized, which leads to an imperfect cancellation of the incoming streak. These characteristics will be further explored in section \ref{understandingtheresults} by means of an evaluation of the SPOD of the actuation effect. In spite of these characteristics, the optimal forcing actuator leads to the lowest actuation energy for the evaluated cases, about one order of magnitude lower than the other two. This is related to the fact that it excites streaks with the highest energy growths and it is therefore capable of leading to a cancellation, specifically at the objective position, with less energy spent.

Finally, the $f_{x_2}$-only actuator presents an intermediary behaviour, leading also to a significant delay in transition. However, it leads to the highest actuation energy.

\section{Discussion}
\label{understandingtheresults}

Two approaches will be used to better understand the results of the previous section: the SPOD of the open and closed-loop cases; and the possibility of causal streak cancellation, which is related to the different delays of the actuators in exciting a response in the output. Both of these analysis are related to the fact that the streamwise velocity fluctuations present their peak value above the wall, as it will be explored next. All results in this section are shown for the $Tu=3.0\%$ case. Trends for $Tu=3.5\%$ are similar and will not be displayed for brevity. 

\subsection{Correlations along the wall-normal direction}

\begin{figure}
\begin{center}
\subfigure[]{\includegraphics[width=0.45\textwidth]{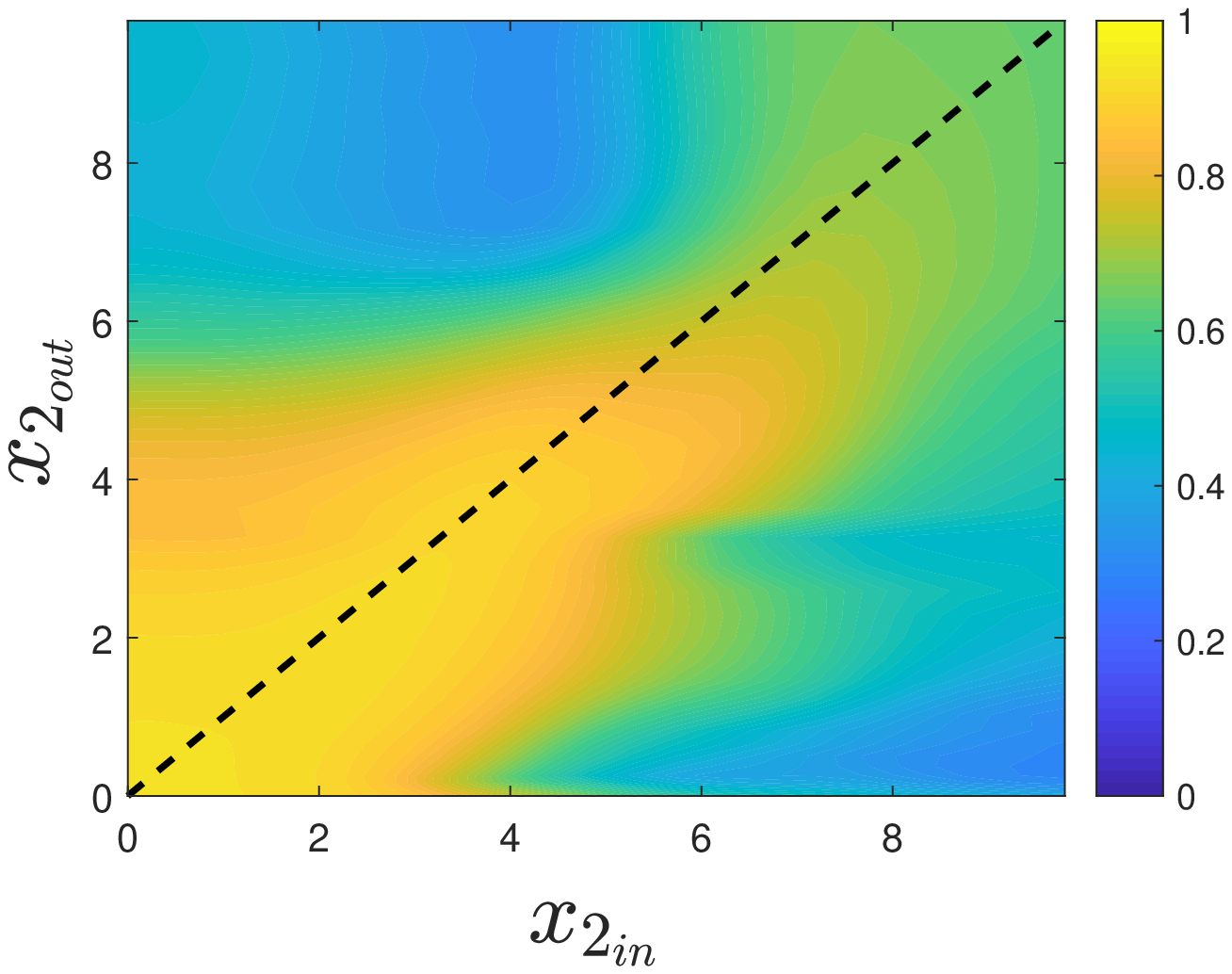}}
\subfigure[]{\includegraphics[width=0.45\textwidth]{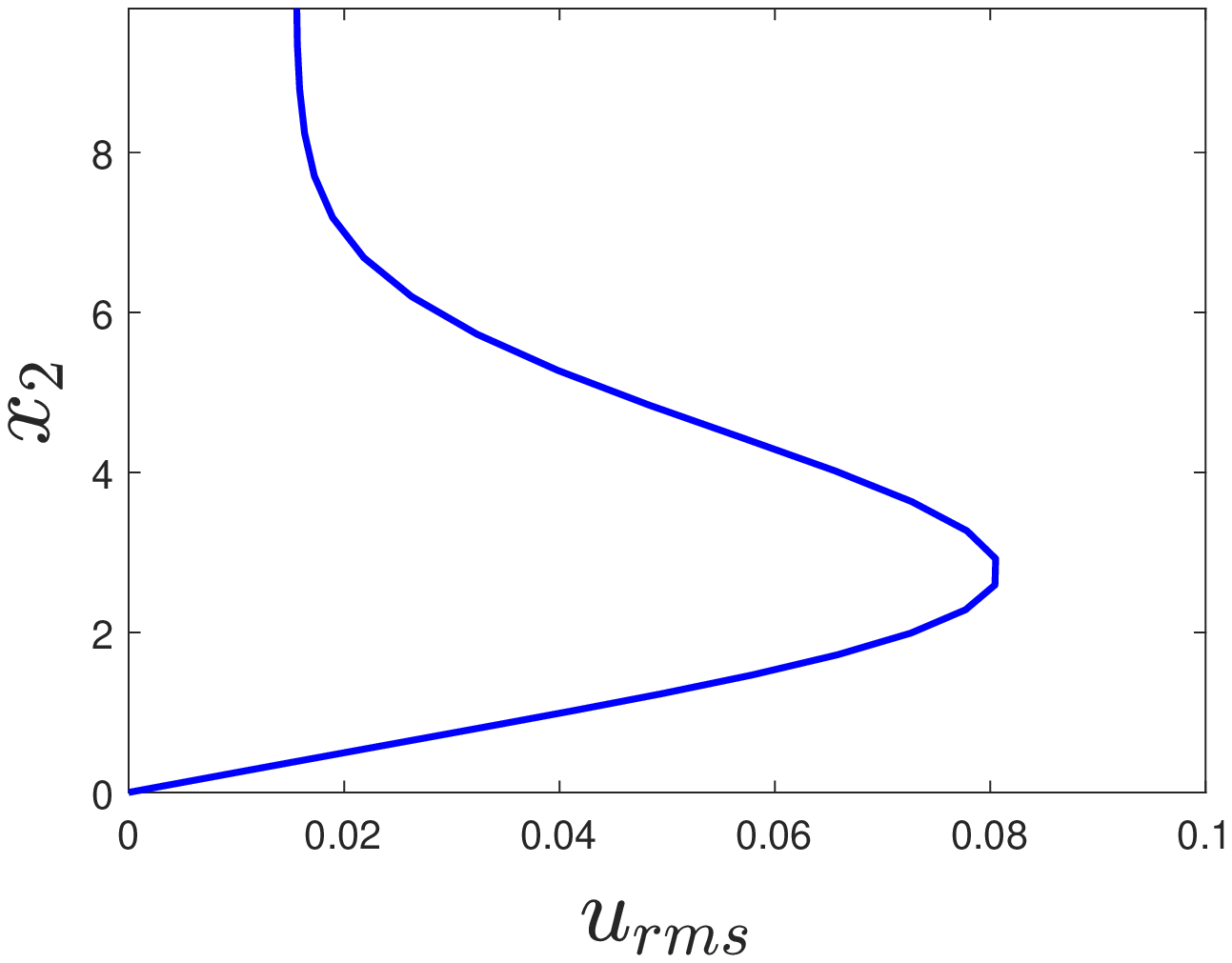}}
\end{center}
   \caption{(a) Correlation between estimated and LES streamwise velocity fields - $x_1$ position of input and output were fixed at 250 and 400, respectively, and the wall-normal position was varied. Dashed line indicates input/output positions at the same wall-normal positions. (b) Root-mean square values of the streamwise velocity fluctuation as a function of the wal-normal direction}
  \label{correlationsandrmsvalues}
\end{figure}

We first consider the RMS value of the streamwise velocity fluctuations along the wall-normal direction, given in figure \ref{correlationsandrmsvalues}. It is noticeable that the maximum value of the RMS is above the wall, located at $x_2 \approx 3$. Therefore, in order to obtain a more global effect on the field the actuator should lead to changes over such higher wall-normal positions rather than in the near-wall region only. Furthermore, the fluctuations at such positions should be predictable given the considered input sensor, which corresponds to measurements of wall shear stress. 

In order to evaluate if the streamwise velocity fluctuations at $x_2=3$ may be predicted from wall-measurements, the correlation between the estimated and LES fields was calculated as a function of the wall-normal positions of input and output measurements. The streamwise positions were kept at $x_1=250$ and 400, for input and output respectively, in accordance with the $y$ and $z$ measurements. The predicted field was obtained by means of the empirical transfer function approach, as outlined in section \ref{flowestcont}. The result is shown in figure \ref{correlationsandrmsvalues} and demonstrates that there is a strong correlation between wall shear stress measurements ($x_{2,in} \to 0$) and the streamwise velocity fluctuations until $x_2 \approx 5$, indicating that the currently considered sensors are adequate for this type of application. Differences are therefore only accountable for the considered actuators, as the method to determine the control law was the same for all of them. In what follows, all the analysis is made with the input sensor corresponding to wall-shear stress.

\subsection{SPOD in open-loop}

\begin{figure}
\centering
\includegraphics[width=0.80\textwidth]{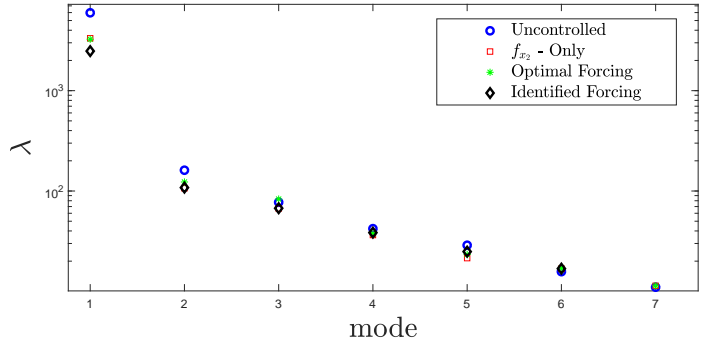}
\caption{Eigenvalues of the SPOD modes calculated at position $x_1=400$, for $(\omega,\beta_k)=(0,0.37)$, for the open (circles) and closed-loop cases (squares, diamonds and crosses, for the blowing, identified and optimal forcing actuator, respectively).}
\label{eigenvaluesofspodforthedifferentcases}
\end{figure}

Spectral POD is calculated at the objective position, $x_1=400$, in the plane defined by the wall-normal and spanwise coordinates. The decomposition is made per $(\omega,\beta_k)$ pair and the results shown here will consider $(\omega,\beta_k)=(0,0.37)$, which corresponds to approximately the most amplified case used in the actuator optimization techniques. The eigenvalues resulting from such decomposition, for the open and closed-loop cases, considering the different actuation strategies evaluated in this paper, are shown in figure \ref{eigenvaluesofspodforthedifferentcases}.

The first mode is approximately two orders of magnitude higher than the second one, a fact that indicates its dominance in the flow, at the evaluated streamwise position. The three evaluated actuators lead to an attenuation of such mode in closed-loop, along with the subsequent modes higher than two. The identified actuator presents the best performance, leading to a reduction of about five times in the magnitude of the first mode eigenvalue. 

\begin{figure}
\centering
\includegraphics[width=0.80\textwidth]{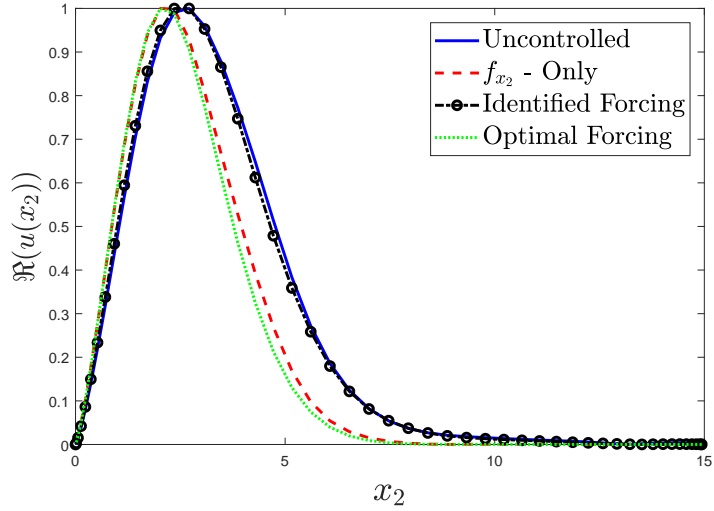}
\caption{First SPOD mode for the uncontrolled (solid line) and different actuation strategies evaluated in this work (dash, dash-dot and dots for the $f_{x_2}$-only, identified and optimal forcing actuator). }
\label{spodforactuationandfstonly}
\end{figure}

In order to better distil the effect of each actuator in the flow, simulations were performed without free-stream turbulence and with the actuators with a white-noise time modulation. SPOD was then applied to the resulting data at the position of objective at the $(\omega,\beta_k)$ pair considered here; the objective is to extract the exact structures that are being excited by the actuators and compare them to the one corresponding to the free-stream-turbulence-induced streaks. Figure \ref{spodforactuationandfstonly} presents the first SPOD mode for the different cases as a function of the wall-normal direction, for the streamwise velocity fluctuation, which is the velocity component that dominates the fluctuation.

It is clear that whereas the identified actuator leads to fluctuations in a compelling agreement with the fluctuations inside the boundary layer, the other two excite ``thinner'' structures with a peak value which occurs closer to the wall than the peak of the streaks inside the boundary layer. This behaviour leads to an imperfect cancellation of the incoming streak. Since a destructive interference is the physical mechanism behind flow control of convectively unstable flows \citep{sasaki2018wave,morra2019_inpress}, the $f_{x_2}$-only and optimal forcing actuators lead to a lower performance in terms of transition delay. It should also be noted that the identified actuator presents a peak in its spatial support which is at a higher wall-normal location than the other two (as shown in figure \ref{forcesblablabla}); this is probably related to the higher peak of the streak it is identifying inside the flow.

%

\subsection{The role of causality in streak cancellation}

\begin{figure}
\begin{center}
\subfigure[Estimated field.]{\includegraphics[width=0.48\textwidth]{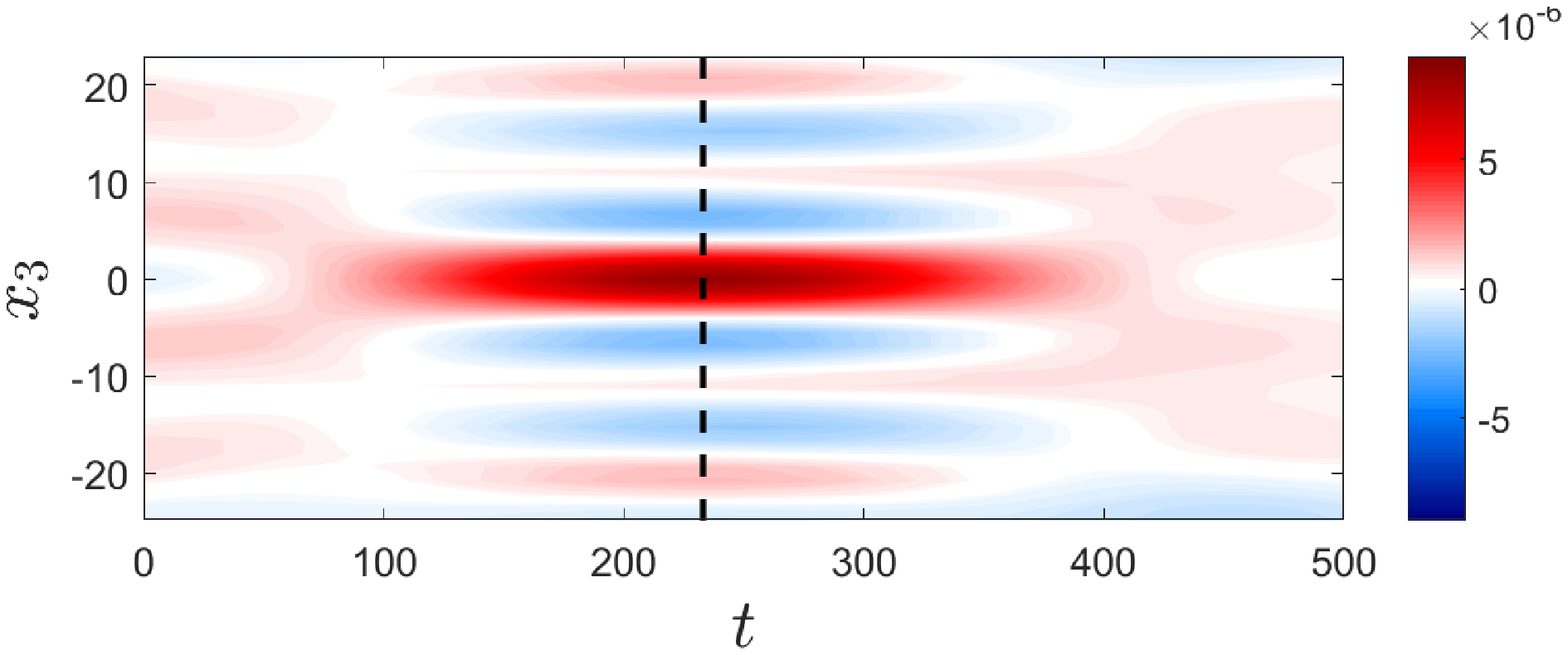}}
\subfigure[Estimated field.]{\includegraphics[width=0.48\textwidth]{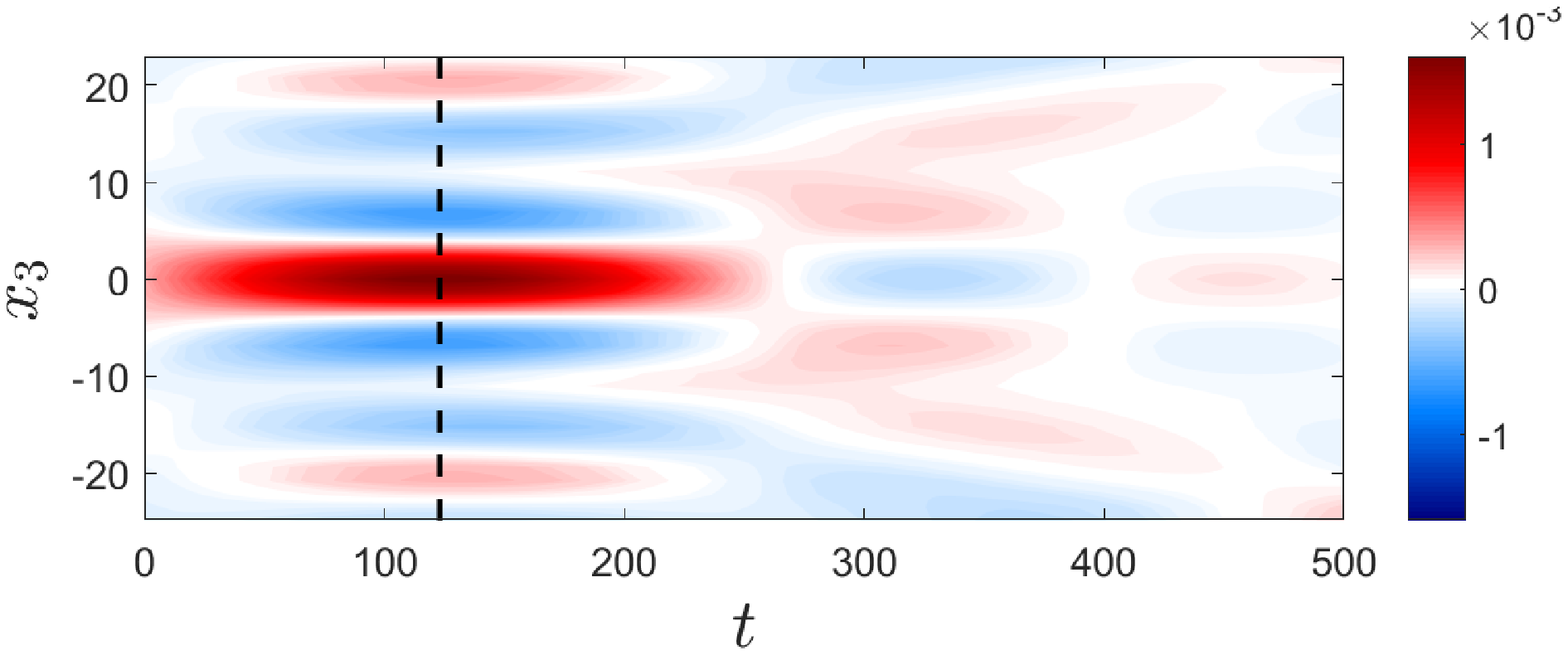}} \\
\subfigure[Optimal forcing actuator.]{\includegraphics[width=0.48\textwidth]{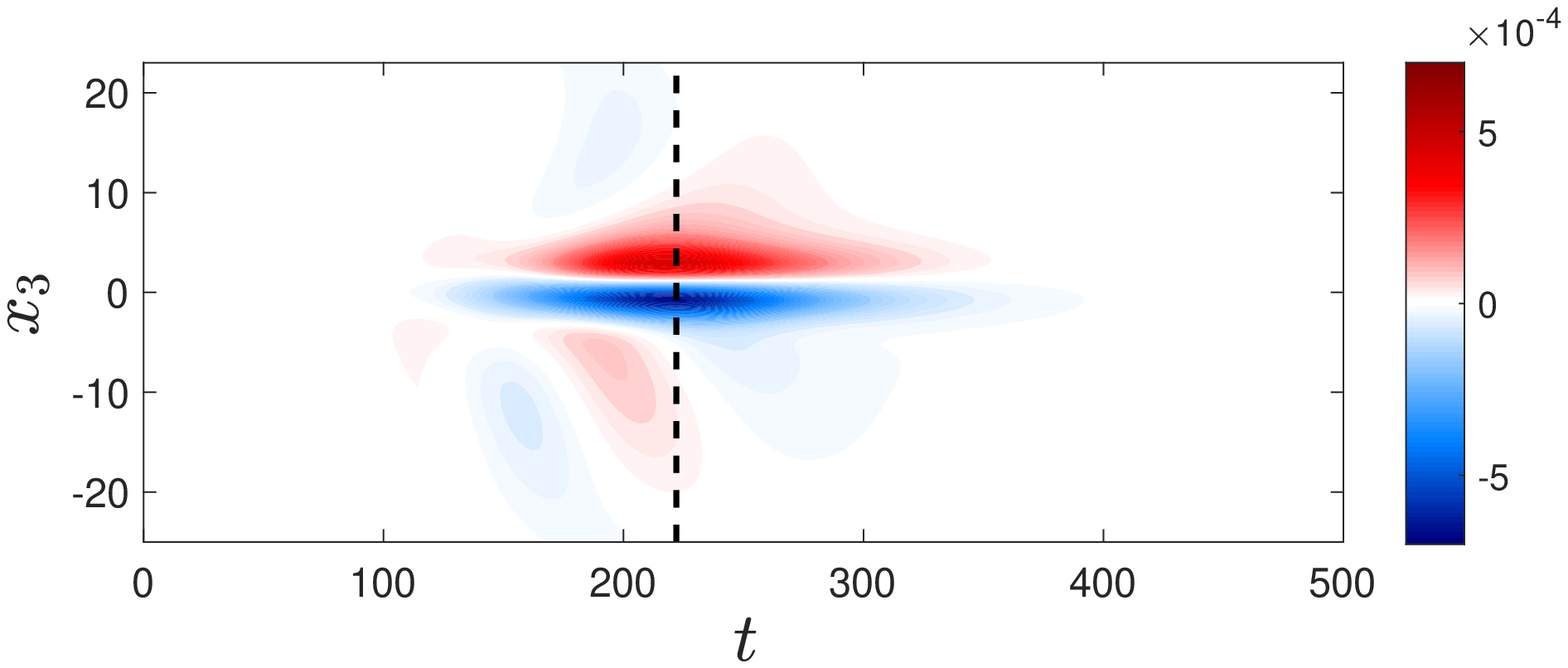}}
\subfigure[Optimal forcing actuator.]{\includegraphics[width=0.48\textwidth]{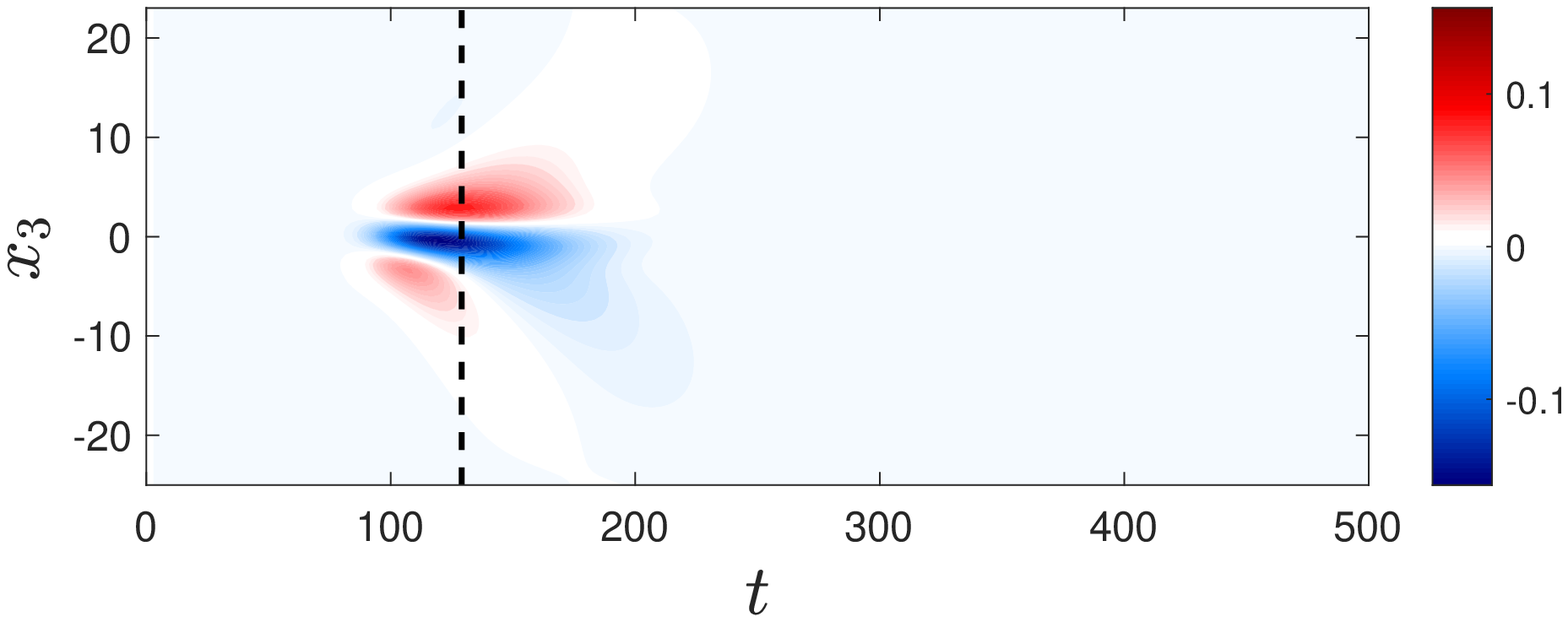}} \\
\subfigure[$f_{x_2}$-only actuator.]{\includegraphics[width=0.48\textwidth]{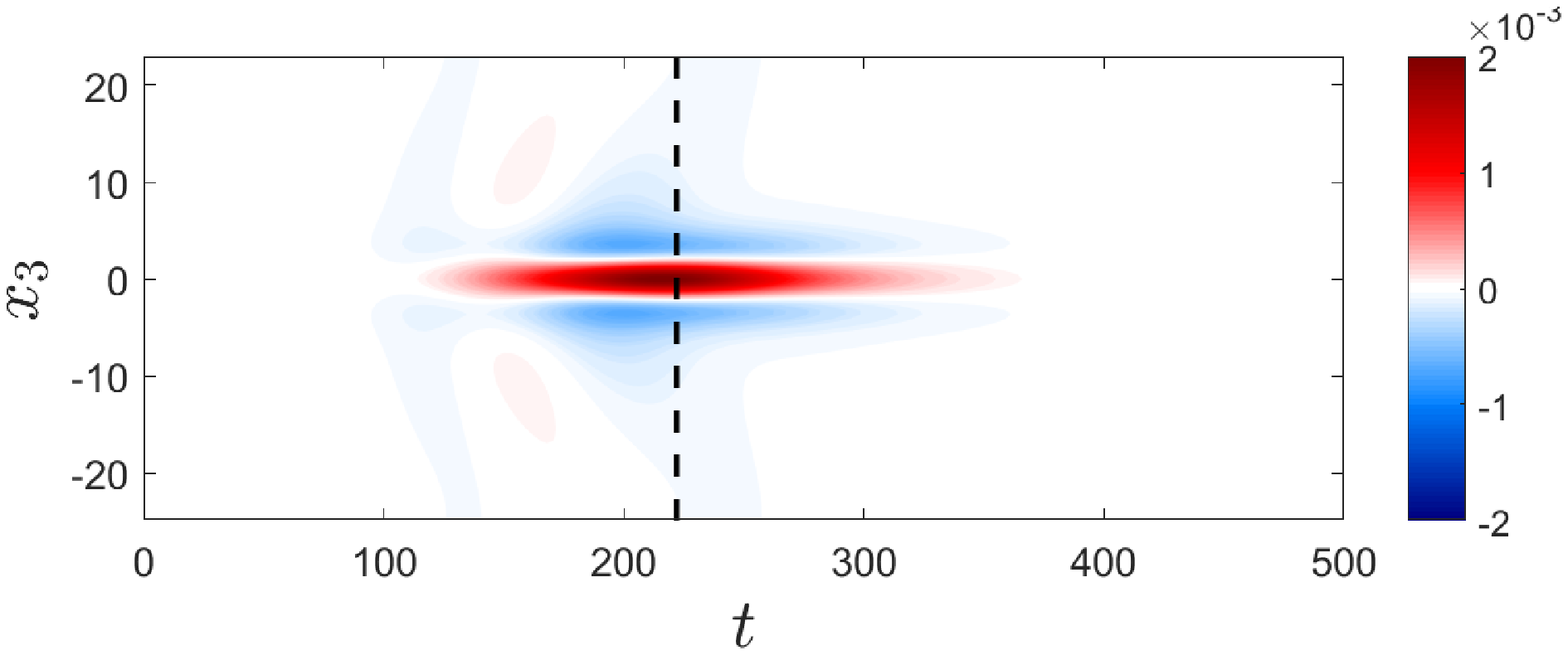}}
\subfigure[$f_{x_2}$-only actuator.]{\includegraphics[width=0.48\textwidth]{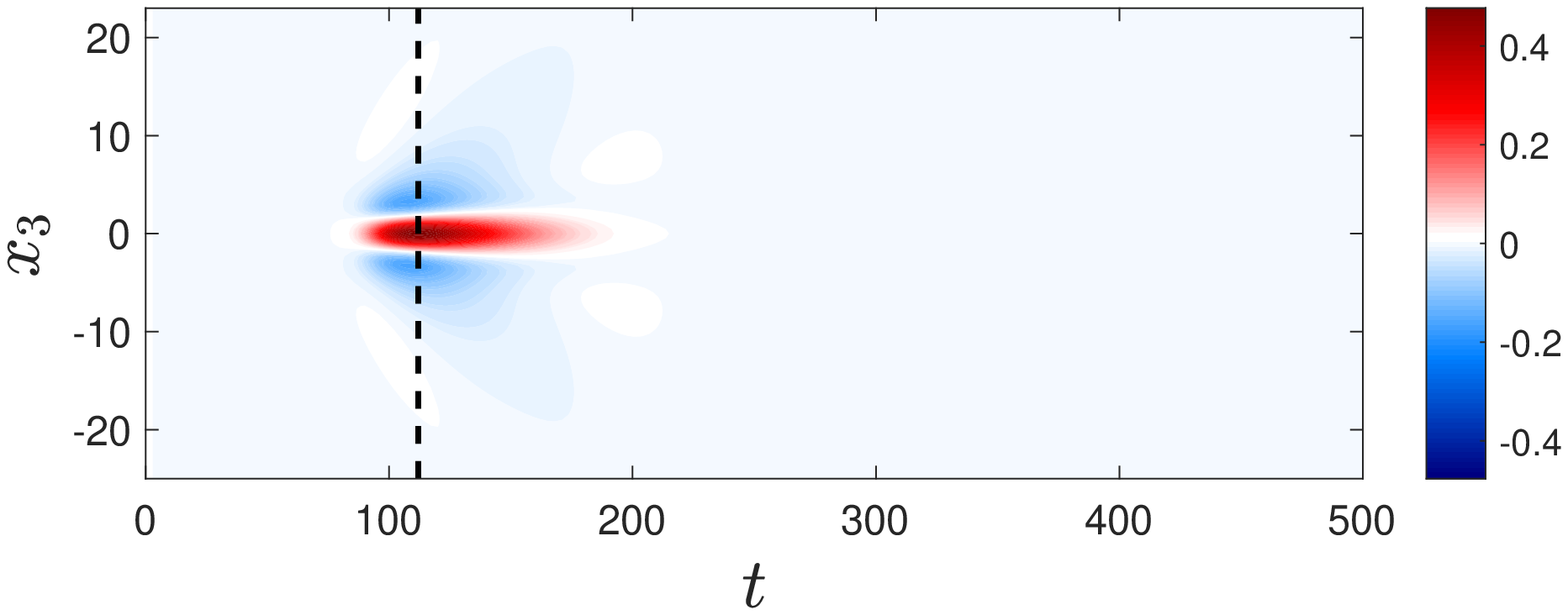}} \\
\subfigure[Identified actuator.]{\includegraphics[width=0.48\textwidth]{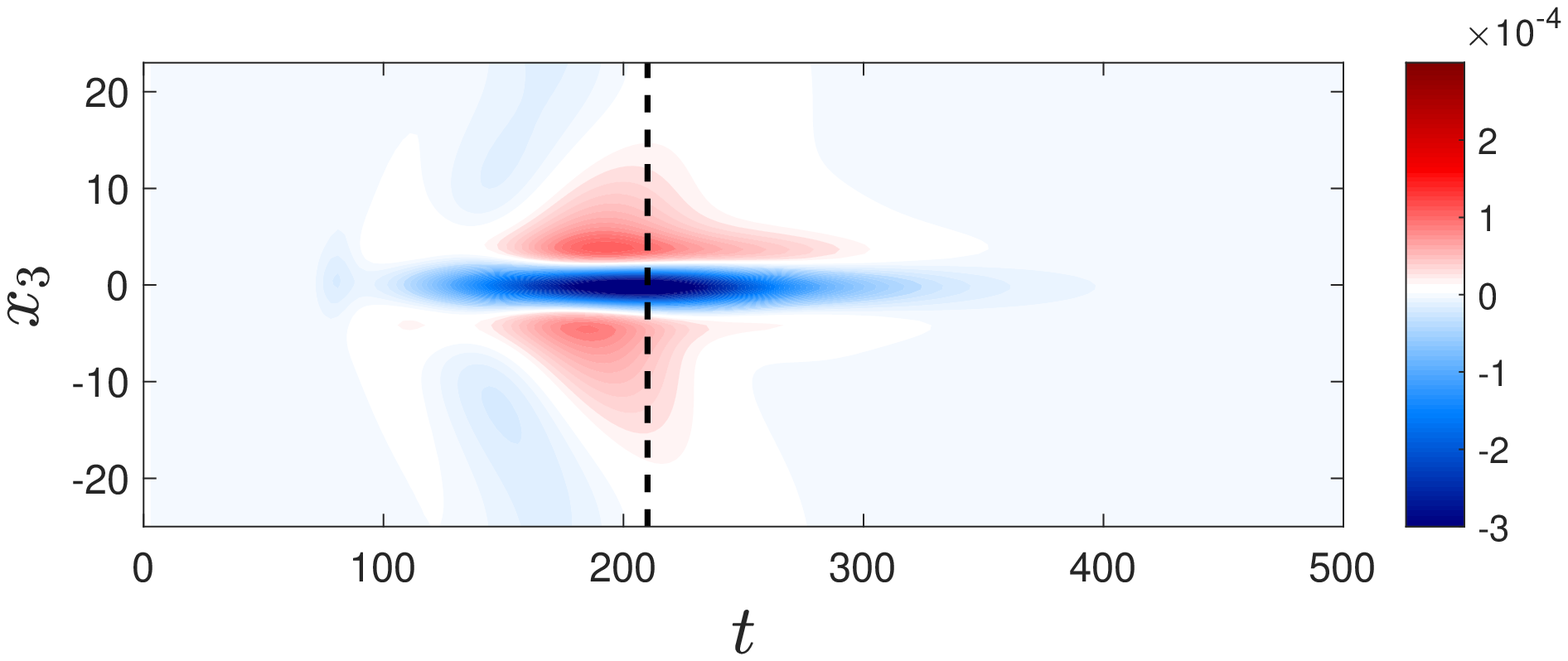}}
\subfigure[Identified actuator.]{\includegraphics[width=0.48\textwidth]{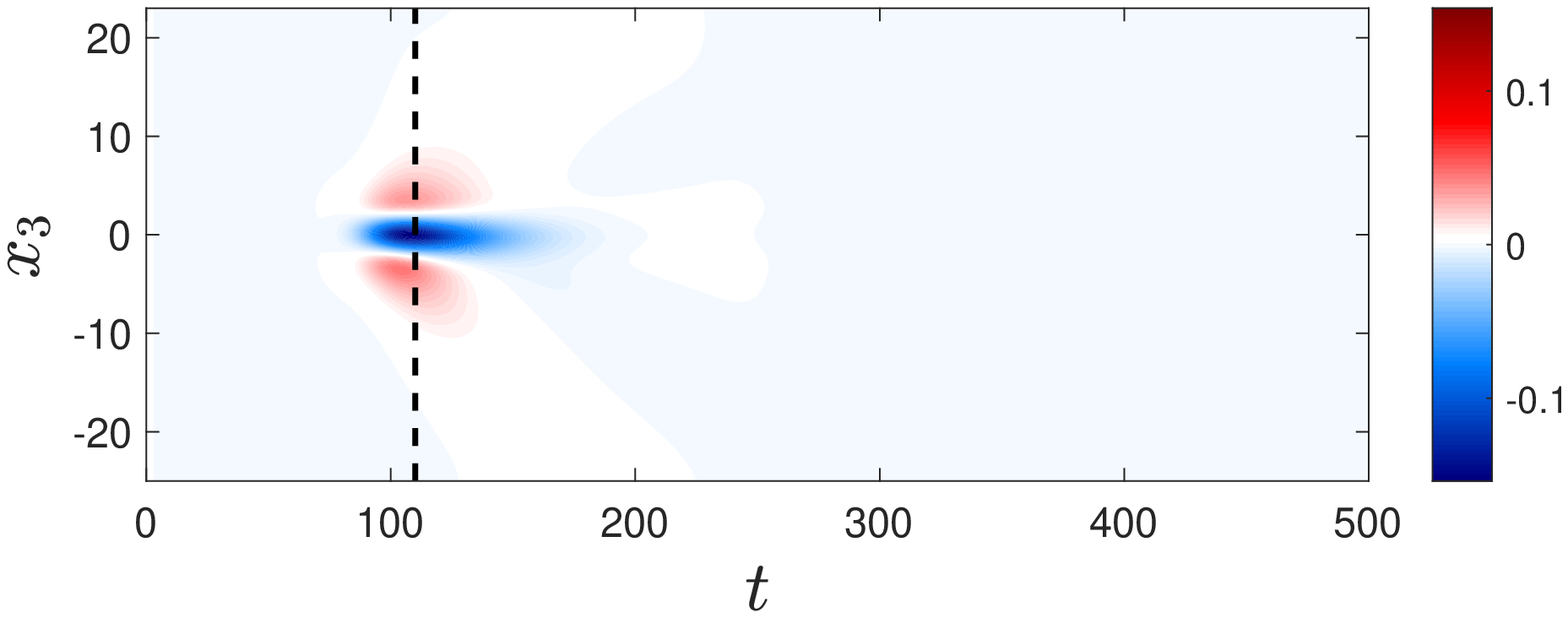}} \\
\end{center}
   \caption{Behaviour of the impulse responses of the estimated field and different actuators considered here. The measurement was located at the wall (left column) and at $x_2=3$ (right column), and the streamwise position was located at $x_1=400$, corresponding to the objective of the control law. The colorbar was adjusted for each plot in order to better visualize the behaviour. The dashed line indicates the time delay for the maximum of the impulse response.}
  \label{impulseresponsesforestimationandactuation}
\end{figure}

Figure \ref{impulseresponsesforestimationandactuation} shows the impulse responses for the estimation (taken between $(x_1,x_2)=(250,0)$, using wall-shear stress as the measurement, and $(400,\epsilon)$ or $(400,3)$, with streamwise velocity as the output measurement, where $\epsilon$ is the first point of the grid above the wall, using the empirical transfer function) and actuation (taken between $(x_1,x_2)=(325,0)$ and $(400,\epsilon)$ or $(400,3)$, considering streamwise velocity as the output measurement). The impulse response for estimation is representative of open-loop disturbances, whereas the one for actuation highlights properties of streaks induced by the control law.

A few characteristics can be observed in figure \ref{impulseresponsesforestimationandactuation}; First, the estimation impulse responses, taken for an input further upstream, present a time delay comparable to the actuation cases, indicating that the group velocity of open-loop streaks is higher than the one corresponding to the actuator-induced disturbances. The delay changes between wall and $x_2=3$ measurements, indicating a tilting of the structures, which reach the higher wall-normal position at $x_1=400$ before reaching the wall at the same streamwise position. Finally, from the time delays of impulse responses it is inferred that the three actuators induce streaks with different group velocity, with the identified case presenting the highest value and optimal forcing the lowest. The relevance of such time delays for closed-loop control are related to the possibility of a causal cancellation of incoming disturbances, as discussed by \cite{sasaki2018_inpress}, and can be summarised, in simplified manner, as follows. Once a given structure is detected by the upstream sensors $y$, it is estimated, through the transfer functions in figure \ref{impulseresponsesforestimationandactuation}, that it will reach the downstream objective $z$ after a time delay $\tau_e$. The actuator should cancel this disturbance, but this cannot occur instantly, since the actuation-induced structures take a time delay $\tau_a$ to reach the objective location. If $\tau_a < \tau_e$ such cancellation is feasible, whereas in the opposite case it becomes impossible to cancel the incoming streaks; in the latter situation, once an upstream streak is detected in $y$, it is already too late to attempt to cancel it in $u$.

Considering the time delays for which the impulse reponse peaks, all actuators are able to generate disturbances that reach the objective position located at the wall before the one related to the estimated field, which can be seen by the characteristic time delays $\tau_a$, lower than $\tau_e$ in figure \ref{impulseresponsesforestimationandactuation} for all cases. The same is not true when we consider the output at $x_2=3$, particularly for the optimal forcing actuator. This characteristic will prevent the optimal forcing case of acting where the highest energy of the streak is present, at the objective position.

\begin{figure}
\begin{center}
\subfigure[Measurement at $x_2=\epsilon$.]{\includegraphics[width=0.90\textwidth]{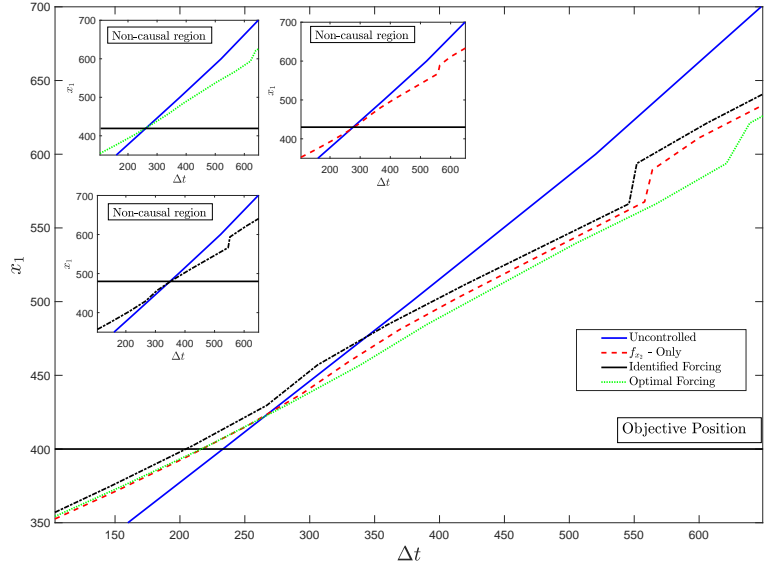}}\\
\subfigure[Measurement at $x_2=3$.]{\includegraphics[width=0.90\textwidth]{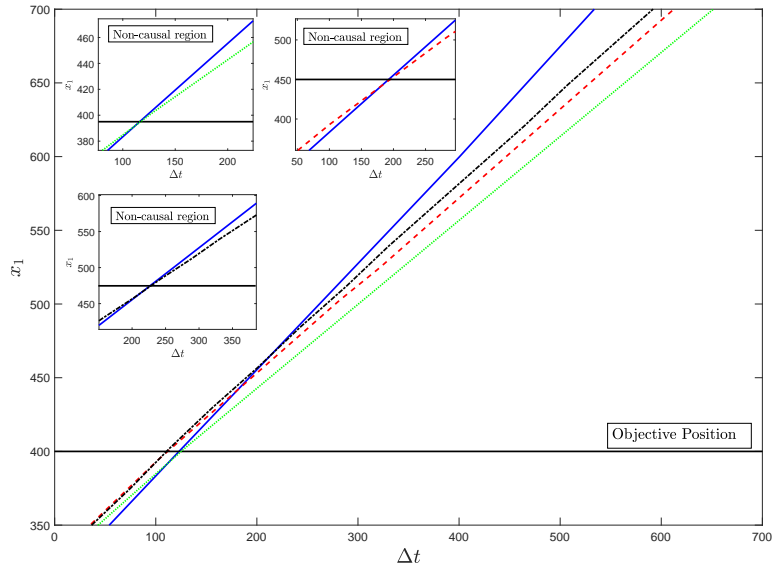}}
\end{center}
   \caption{Streamwise position for the different actuators and estimated field as a function of time. Input positions ($\Delta t =0$) were kept fixed at $(x_1,x_2)=(250,0)$, with wall-shear stress, and $(x_1,x_2)=(325,0)$, using streamwise velocity, for estimation and actuation cases, respectively. The transverse position of the measured impulse corresponds to (a) $x_2=\epsilon$ and (b) $x_2=3$.}
  \label{delaysforthedifferentactuators}
\end{figure}

This characteristics are summarized in figure \ref{delaysforthedifferentactuators}, where the streamwise position is given as a function of the time-delay for it to be reached for the estimation and each actuator. The delay was considered as the time when the peak value is reached, for each impulse response at the considered position. The slope of lines in the plot can thus be related to the group velocity of disturbances in the open-loop case, given by the estimation transfer function, and of the ones resulting from the three actuators. The values at the wall (considered as $x_2=\epsilon$) and at $x_2=3$ are shown in figure \ref{delaysforthedifferentactuators}. The input measurements were considered as $x_1=250$, for estimation, and $x_2=325$, for actuation. For all actuators and both wall-normal positions it is noticeable that the impulse response of the estimation has a higher velocity. Among the three actuators, the identified one leads to structures with higher group velocity, which is particularly clear for the $x_2=3$ case, the most important one in terms of the energy content of the fluctuations. This higher velocity can be related to generation of streaks at higher wall-normal positions (figure \ref{spodforactuationandfstonly}), and may further justify the effectiveness of the identified actuator.

It should be noted that when the curve corresponding to the estimated field surpasses those of the impulses, these positions correspond to uncontrollable cases, as the control-induced streaks will reach a downstream position after the incoming structures one wishes to attenuate. The considered objective position of $x_1=400$ is highlighted and the impulse responses of all actuators reach it before the estimated field, which therefore leads to a causal kernel which is capable of the attenuations reported here, which result in a good attenuation of the objective quantity.

However, when one considers the $x_2=3$ case, where most of the fluctuation energy is contained, the streamwise objective position of $x_1=400$ does not correspond to a causal behaviour when the optimal forcing actuator is considered, which explains why it presents the worst performance among the different spatial supports considered here. The higher group velocity of streaks for larger $x_2$  prevents a perfect cancellation particularly for the optimal forcing and vertical forcing-only actuators, which present considerably lower values of group velocity.

One could try to compensate for the lower group velocity of the actuator-induced streaks by moving the objective position upstream, closer to the actuator. This would cause all schemes to be causal, both at wall and at $x_2=3$. However, since the free-stream turbulence is continuously forcing the boundary layer, it is expected that the lower group velocity of the actuator-induced structures, particularly for the vertical-force and optimal cases, will eventually play a role in downstream areas of the flow. 

\section{Conclusions}
\label{concludingthepaper}

Three methodologies have been considered for the design of localized actuators for the control of streaky structures, induced by free-stream turbulence. A set-up close to practical applications was considered, where a large number of OSS modes was used to model free-stream turbulence. This makes it infeasible to compute individual impulse responses of each disturbance, and an empirical transfer function was derived to obtain a reduced-order model, for which application of LQG led to control laws. 

Two of the methods of actuator design corresponded to optimization procedures, where either the energy of the actuator induced disturbances at the position of objective, or the difference with respect to a previously measured structure between actuator induced and open-loop disturbances was considered as the cost function, which was maximized in the first case (leading to an optimal forcing) and minimized in the second (with an identified, tailored actuator that optimally target open-loop streaks). The resulting direct/adjoint iteration algorithm was computationally efficient and led to desired results when applied in the nonlinear simulation. A third actuator, corresponding to a vertical forcing-only served as a baseline case, in line with the recent results of \cite{shahriari2018control} where a ring of plasma actuators was used to excite a vertical body force in a boundary layer.

Closed-loop control with all the actuators led to significant delay in transition, and this was shown to be robust to mild changes in the free-stream turbulence level, a desired characteristic for real-life applications. Differences between the three cases were understood in terms of the SPOD of estimation and actuation fields, in open-loop, which highlighted the dissimilarities between the structures induced by the actuators and the one actually present in a boundary layer. Here, an important difference between control of TS waves and streaks appears. Whereas in the former case any actuator leads to exactly the same TS waves at downstream positions, as these are the only structures in spatial growth, for streaks a whole family of disturbances can be generated by actuators. It thus becomes important to target precisely the streaks that are actually expected in a given transitional boundary layer, and thus the identified actuator obtains a closed-loop performance superior to the other ones, since it cancels more accurately the open-loop streaks.

The distinct velocities of structures induced by the three actuators and the streaks induced by the free-stream turbulence, along with the tilting of the structures along the wall-normal direction, also plays a role in this type of application, where it may become impossible to obtain a causal cancellation of incoming disturbances, even if the actuator is downstream of the input measurement. Causality will also depend on the wall-normal position under consideration, a feature which had not yet been observed or quantitatively computed.

Finally, an evaluation of the correlations along the wall-normal direction indicated that wall measurements were adequate for the prediction of the output signals. Such technique allowed similar conclusions to observability tools without the need to perform adjoint simulations. These analysis were possible by means of the empirically calculated impulse responses, which permitted an exploration of the parameters of the problem reducing the number of required non-linear simulations.

On what concerns the design of actuators for experimental applications, it was shown that a vertical forcing only, which is currently possible to implement, should be adequate for the control of streaky structures. Better results may be obtained both in terms of the transition delay and energy budget of actuation when access to the open-loop data is available prior to the design of the actuator, where the methods outlined here for the evaluation of the forcing and optimizations should aid in the design of new actuators for flow control.

\bigskip


The authors would like to acknowledge the VINNOVA Projects PreLaFlowDes and SWE-DEMO and the Swedish-Brazilian Research and Innovation Centre CISB for funding. Moreover, part of this work was performed during an exchange programme at KTH, for which Kenzo Sasaki received a scholarship from Capes, project number 88881.132008/2016-01. Kenzo Sasaki is also funded by a scholarship from FAPESP, grant number 2016/25187-4. Andr\'e V. G. Cavalieri was supported by a CNPq grant 310523/2017-6. The simulations were performed on resources provided by the Swedish National Infrastructure for Computing (SNIC) at NSC, HPC2N and PDC.

\appendix

\section{Calculation of SPOD modes from data}

We briefly outline the approach to compute SPOD modes from snapshots taken from a simulation or experimental data. As in the main body of the paper, $\mathbf{q}(x,y,z,t)=(u(x_1,x_2,x_3,t),v(x_1,x_2,x_3,t),w(x_1,x_2,x_3,t))$, which is then Fourier transformed from $x_3$ to $\beta_k$,

\begin{equation}
\label{eqspodfromdata1}
\hat{\mathbf{q}}(x_1,x_2,\beta,t)=\int_{-x_{2_{max}}/2}^{x_{2_{max}}/2} \mathbf{q}(x_1,x_2,x_3,t) e^{-i\beta \zeta}\mathrm{d}\zeta.
\end{equation}

\noindent The spanwise direction is discretized and the continuous Fourier transform becomes a discrete Fourier transform (DFT), which is evaluated at the discrete wavenumbers $\beta_k$. Given the periodicity of this coordinate, the DFT is regarded as the discretization of the coefficients of the corresponding Fourier series.

Consider now the discretization in time, the quantity $\hat{\mathbf{q}}_k(\beta_k)$ represents the instantaneous state of $\hat{\mathbf{q}}(\beta_k,t)$. If a total number of $N$ snapshots is used, the signal may be regarded as:

\begin{equation}
\label{eqspodfromdata2}
\hat{\mathbf{Q}}(\beta_k)=\left[\hat{\mathbf{q}}_1(\beta_k) \quad \hat{\mathbf{q}}_2(\beta_k) \quad \cdots \quad \hat{\mathbf{q}}_N(\beta_k) \right],
\end{equation}

\noindent where $\hat{\mathbf{Q}}(\beta_k)$ is $N_s \times N$, $N_s$ representing the number of spatial grid points times the number of physical quantities considered (on this case, the three velocity components and pressure). Application of the DFT directly into the lines of matrix $\hat{\mathbf{Q}}(\beta_k)$ should not be performed as the result will not converge with the number of snapshots \citep{bendat2011random}, and the order of magnitude of the error could be as high as the corresponding magnitude of the spectrum. Therefore, in order to obtain converged values of the spectral density, for calculation of the spectral density tensor, it is necessary to average the spectra over multiple realizations of the flow. This may be accomplished by application of Welch's method \citep{welch1967use}.

Start by partitioning the full signal into $N_b$ blocks, each with $N_f$ elements, the nth block is then given as:

\begin{equation}
\label{eqspodfromdata3}
\hat{\mathbf{Q}}^{(n)}(\beta_k)=\left[\hat{\mathbf{q}}^{(n)}_1(\beta_k) \quad \hat{\mathbf{q}}^{(n)}_2(\beta_k) \quad \cdots \quad \hat{\mathbf{q}}^{(n)}_{N_f}(\beta_k) \right],
\end{equation}

\noindent such that each block can be regarded as a realization of the flow. Overlapping the blocks with adjacent elements is possible and allows a higher number of blocks for the same length of the original signal, permitting a faster convergence of the statistics. The $k$th entry in the $n$th block is then given as $\hat{\mathbf{q}}^{(n)}_k(\beta_k)=\hat{\mathbf{q}}_{{k+(n-1)(N_f-N_o)}}(\beta_k)$, where $N_o$ is the number of overlapping snapshots. The DFT is then calculated for at each block,

\begin{equation}
\label{eqspodfromdata4}
\hat{\hat{\mathbf{Q}}}^{(n)}(\beta_k)=\left[\hat{\hat{\mathbf{q}}}^{(n)}_1(\beta_k) \quad \hat{\hat{\mathbf{q}}}^{(n)}_2(\beta_k) \quad \cdots \quad \hat{\hat{\mathbf{q}}}^{(n)}_{N_f}(\beta_k) \right],
\end{equation}

\noindent where the $k$th element of the block is then given as:

\begin{equation}
\label{eqspodfromdata5}
\hat{\hat{\mathbf{q}}}^{(n)}_{k}(\beta_k)=\frac{1}{\sqrt{N_f}}\sum_{j=1}^{N_f}w_j\hat{\hat{\mathbf{q}}}^{(n)}_{j}(\beta_k)e^{-2\pi i (k-1)[(j-1)/N_f]},
\end{equation}

\noindent with $k=1, \cdots ,N_f$ and $n=1,\cdots , N_b$. Equation \eqref{eqspodfromdata5} represents the discrete Fourier transform of the block, with the addition of the weights $w_j$ which allow the application of a window function, used to reduce spectral leakage due to the non-periodicity of the block. The normalization factor $1/N_f$ ensures the transform is unitary for a square window. $\hat{\hat{\mathbf{q}}}^{(n)}_{k}(\beta_k)$ is the kth element of the DFT of the nth block, with a corresponding frequency $\omega_k$, 

\begin{equation}
\label{eqspodfromdata6}
\omega_k=2\pi\frac{k-1}{n\Delta T}, k\leq n/2
\end{equation}

\noindent or

\begin{equation}
\label{eqspodfromdata7}
\omega_k=2\pi\frac{k-1-n}{n\Delta t}, k > n/2.
\end{equation}

Finally, the cross-spectral density tensor $\mathbf{R}(\mathbf{x},\mathbf{x}^\prime,\omega,\beta_k)$ can be estimated at a frequency $\omega_k$ and spanwise wavenumber $\beta_k$ by averaging the blocks,

\begin{equation}
\label{eqspodfromdata8}
\mathbf{R}_{\omega_k}(\beta_k)=\frac{\Delta t}{s N_b}\sum_{n=1}^{N_b}\hat{\hat{\mathbf{q}}}^{(n)}_k(\beta_k)\left(\hat{\hat{\mathbf{q}}}^{(n)}_k(\beta_k)\right)^*,
\end{equation}

\noindent and $s=\sum_{j=1}^{N_f}w_j^2$. Each Fourier coefficient at a frequency $\omega_k$ for each block, at a given $\beta_k$, can be arranged in a matrix form,

\begin{equation}
\label{eqspodfromdata9}
\hat{\hat{\mathbf{Q}}}_{\omega_k}(\beta_k)=\sqrt{k}\left[\hat{\hat{\mathbf{q}}}^{(1)}_k(\beta_k) \quad \hat{\hat{\mathbf{q}}}^{(2)}_k(\beta_k) \quad \cdots \quad \hat{\hat{\mathbf{q}}}^{(N_b)}_{k}(\beta_k)\right]
\end{equation}

\noindent where $k=\Delta t/(sN_b)$ and $\hat{\hat{\mathbf{Q}}}_{\omega_k}(\beta_k)$ is $N \times N_b$. The cross-spectral density tensor is then written compactly as,

\begin{equation}
\label{eqspodfromdata10}
\mathbf{R}_{\omega_k}(\beta_k)=\hat{\hat{\mathbf{Q}}}_{\omega_k}(\beta_k)\left(\hat{\hat{\mathbf{Q}}}_{\omega_k}(\beta_k)\right)^*.
\end{equation}

The calculation of the cross-spectral density tensor then converges as the number of blocks and snapshots at each block is increased together \citep{bendat2011random}.

Defining the positive-define Hermitian matrix $\mathbf{W}$, $N \times N$, to account for the weight and the numerical quadrature for performing an integral on a discrete grid, the SPOD eigenvalue problem reduces to an $N \times N$ matrix eigenvalue problem, at each frequency and transverse wavenumber,

\begin{equation}
\label{eqspodfromdata11}
\mathbf{R}_{\omega_k}(\beta_k)\mathbf{W}\psi_{\omega_k}(\beta_k)=\psi_{\omega_k}(\beta_k)\lambda_{\omega_k}(\beta_k).
\end{equation}

The SPOD modes are then given in the columns of $\psi_{\omega_k}(\beta_k)$, ranked accordingly to their corresponding eigenvalues, which are in the diagonal matrix $\psi_{\omega_k}(\beta_k)$. 

\section{Derivation of the Optimization Schemes}
\label{appendixwiththeadjointoptimization}

In this section, the two schemes for the calculation of the actuators considered in subsections \ref{optimalforcingactuator} and \ref{identifiedactuator} will be outlined. We follow the works of \cite{andersson1999optimal} and \cite{levin2003exponential} in a scheme appropriate for algebraically growing disturbances along the streamwise direction, for a slowly divergent mean flow. The pressure and velocity fluctuations follow the boundary layer equations which, written in matrix form, are given as:

\begin{equation}
\label{blequations1}
\mathbf{\mathcal{A}}\mathbf{\hat{q}}+\mathbf{\mathcal{B}}\frac{\partial\mathbf{\hat{q}}}{\partial x_2}+\mathbf{\mathcal{C}}\frac{\partial^2\mathbf{\hat{q}}}{\partial x_2^2}+\mathbf{\mathcal{D}}\frac{\partial\mathbf{\hat{q}}}{\partial x_1}=\mathbf{\hat{F}}
\end{equation}

\noindent where $\mathbf{\hat{F}}=(0,\hat{F}_{x_1},\hat{F}_{x_2},\hat{F}_{x_3})$ is a forcing applied in the three directions and $\mathbf{\hat{q}}=(\hat{u},\hat{v},\hat{w},\hat{p})$, and the hat indicates quantities given in the $(\omega,\beta)$ domain. Equation \eqref{blequations1} results from the application of the Ansatz $\mathbf{{q}}=\mathbf{\hat{q}}(x,y)e^{(i\beta z-i \omega_t)}$ into the linearized Navier--Stokes equations. The operators are then given as:

\begin{equation}
\mathbf{\mathcal{A}}=
\begin{pmatrix}
0                                           & 0                                          & i\beta                   & 0\\
-i\omega+\beta^2/Re+\mathrm{d}U/\mathrm{d}x_1 & \mathrm{d}U/\mathrm{d}x_2                    & 0                        & 0\\
0                                           & -i\omega+\beta^2/Re+\mathrm{d}V/\mathrm{d}x_2& 0                        & 0\\
0                                           & 0                                          & -i\omega+\beta^2/Re      & i\beta\\
\end{pmatrix}
\end{equation}

\begin{equation}
\mathbf{\mathcal{B}}=
\begin{pmatrix}
0 & I & 0 & 0\\
V & 0 & 0 & 0\\
0 & V & 0 & I\\
0 & 0 & V & I\\
\end{pmatrix}
\end{equation}

\begin{equation}
\mathbf{\mathcal{C}}=
\begin{pmatrix}
0     & 0     & 0     & 0\\
-I/Re & 0     & 0     & 0\\
0     & -I/Re & 0     & 0\\
0     & 0     & -I/Re & 0\\
\end{pmatrix}
\end{equation}

\begin{equation}
\mathbf{\mathcal{D}}=
\begin{pmatrix}
I & 0 & 0 & 0\\
U & 0 & 0 & 0\\
0 & U & 0 & 0\\
0 & 0 & U & 0\\
\end{pmatrix}
\end{equation}

\noindent where $U$ and $V$ are the mean velocity components in the streamwise and wall-normal direction, respectively and $I$ is the identity matrix. The same non-dimensionalizations as in the remaining of the paper are considered here. It should also be noted that the streamwise wavenumber, normally referred to as $\alpha$, is not present in this equations as there will be no exponential dependence of the fluctuations and all streamwise variation is to be absorbed in $\mathbf{\hat{q}}(x_1,x_2,\omega,\beta)$, which implies that Tollmien-Schlichting waves will not be considered here. This equations are appropriate for algebraically growing disturbances, as the streaky structures induced by the free-stream turbulence.

Equation \ref{blequations1} can be written in compact form as

\begin{equation}
\label{onemoreequation}
\mathcal{L}\hat{\mathbf{q}}=\hat{\mathbf{F}}
\end{equation}

\noindent  where the spatial, frequency and wavenumber dependences have been absorbed into $\mathcal{L}$ as 

\begin{equation}
\mathcal{L}=\mathbf{\mathcal{A}}+\mathbf{\mathcal{B}}D_{x_2}+\mathbf{\mathcal{C}}D_{x_2}^2+\mathbf{\mathcal{D}}D_{x_1}
\end{equation}

\noindent where the derivative operators $D_{x_2}$ and $D_{x_1}$ represent discretized derivative operations in the wall-normal and streamwise directions, respectively.

Equation \eqref{blequations1} is integrated in the streamwise direction using a first or second order explicit Euler method. We solve the problem subject to an initial condition at $x_1 = 0$, and, since the equation is parabolic, downstream spatial marching can be performed. The discretization over the wall-normal direction is made by means of Chebyshev polynomials considering 300 points. Dirichlet boundary conditions are applied to the velocity components at the wall and at $x_2 \rightarrow \infty$.

The objective is to minimize a cost function which considers the difference between the calculated fluctuation, at the objective position, and the SPOD of the field in open-loop. A similar approach has been used by \cite{tissot2017sensitivity} to identify forcing terms in a turbulent jet. The considered cost-function is then:

\begin{equation}
\label{costfunction}
E_f=\frac{1}{2}\int_0^\infty \left.\|\hat{\mathbf{q}}-\hat{\mathbf{q}}_{SPOD}\|^2 \right\rvert_{x_1=x_{1_f}}\mathrm{d}x_2
\end{equation}

\noindent it should be noted that by considering $\hat{\mathbf{q}}_{SPOD}=0$ and performing a maximization, rather than a minimization, the usual procedure to obtain the optimal forcing is recovered. This will be treated as a particular case of this procedure.

We then follow the approach of \cite{pralits2000sensitivity} by defining an extended Lagragian functional which includes the cost function and the constraint, which is given in terms of equation \ref{onemoreequation}, with the addition of a Lagrange multiplier, which plays the role of the adjoint variable, $\hat{\mathbf{q}}^*$.

\begin{equation}
\label{extendedfunctional}
J = E_f-Re(\langle \hat{\mathbf{q}}^*,\mathcal{L}\hat{\mathbf{q}}-\hat{\mathbf{F}} \rangle)
\end{equation}

\noindent The brackets represent the inner product, which is defined for two arbitrary functions as:

\begin{equation}
\label{innerproductdefinition}
\langle \phi, \psi \rangle = \int_{x_{1_0}}^{x_{1_f}} \int_0^{\infty} \overline{\phi}(x_1,x_2)\psi (x_1,x_2) \mathrm{d}x_2 \mathrm{d}x_1
\end{equation}

We then take the variation of equation \ref{extendedfunctional}, which leads to:

\begin{equation}
\label{fdsalkfjadj}
\delta J = \int_0^\infty \left.(\overline{\hat{\mathbf{q}}}-\overline{\hat{\mathbf{q}}_{SPOD}})\right\rvert_{x_1=x_{1_f}}\delta \hat{\mathbf{q}} \mathrm{d}x_2 -\langle \hat{\mathbf{q}}^*,\mathcal{L}\delta \hat{\mathbf{q}}-\delta\hat{\mathbf{F}} \rangle-\langle \delta\hat{\mathbf{q}}^*,\mathcal{L} \hat{\mathbf{q}}-\hat{\mathbf{F}} \rangle
\end{equation}

\noindent the sensitivity of the Lagrangian functional to infinitesimal changes $\delta \mathbf{\hat{q}}$,   $\delta \mathbf{\hat{F.}}$ and $\delta \mathbf{\hat{q}^*}$. When the variation becomes zero, the Lagrangian is minimized or maximized. In this case, the third term $\langle \delta\hat{\mathbf{q}}^*,\mathcal{L} \hat{\mathbf{q}}-\hat{\mathbf{F}} \rangle$ is equal to zero, such that the state equation in \eqref{onemoreequation} is satisfied, as desired. Zeroing the second term will result in the adjoint problem and corresponding boundary and initial conditions, as follows. 

The operators are moved to the right side of the inner product by considering the following property of the adjoint,

\begin{equation}
\langle \hat{\mathbf{q}}^*,\mathbf{\mathcal{A}} \hat{\mathbf{q}} \rangle=\langle \mathbf{\mathcal{A}}^*\hat{\mathbf{q}}^*,\hat{\mathbf{q}} \rangle
\end{equation}

\noindent where the star $*$, refers to a conjugate transpose, when applied to a matrix operator. The derivatives are moved from the direct to the adjoint problem by integrations by parts. We then obtain:

\begin{equation}
\langle \hat{\mathbf{q}}^*,\mathcal{L}\delta \hat{\mathbf{q}}-\delta\hat{\mathbf{F}} \rangle = \langle (\mathbf{\mathcal{A}}^*-\mathbf{\mathcal{B}}^*_{x_2}-\mathbf{\mathcal{D}}^*_{x_1})\hat{\mathbf{q}}^*-\mathbf{\mathcal{B}}^*\hat{\mathbf{q}}^*_{x_2}+\mathbf{\mathcal{C}}^*\hat{\mathbf{q}}^*_{x_2x_2}-\mathbf{\mathcal{D}}^*\hat{\mathbf{q}}^*_{x_1},\hat{\mathbf{q}} \rangle -\langle \hat{\mathbf{q}}^*,\delta \hat{\mathbf{F}\rangle}+b.c.
\end{equation}

\noindent where the subscripts $x_1$ and $x_2$ indicate that the corresponding operator has been derived with respect to $x_1$ or $x_2$. Setting this to zero for arbitrary $\delta\mathbf{\hat{q}}$ leads to the adjoint boundary layer equations, given by

\begin{equation}
\label{lkfjaslfjsald}
(\mathbf{\mathcal{A}}^*-\mathbf{\mathcal{B}}^*_{x_2}-\mathbf{\mathcal{D}}^*_{x_1})\hat{\mathbf{q}}^*-\mathbf{\mathcal{B}}^*\hat{\mathbf{q}}^*_{x_2}+\mathbf{\mathcal{C}}^*\hat{\mathbf{q}}^*_{x_2x_2}-\mathbf{\mathcal{D}}^*\hat{\mathbf{q}}^*_{x_1}=0
\end{equation}

\noindent where the subscripts $x_1$ and $x_2$ represent derivatives along the corresponding directions.

The term $b.c.$ corresponds to four integrals. Once set to zero, it will supply the boundary conditions for the adjoint boundary layer equations. Writing explicitly the first three, we have:

\begin{equation}
\int_{x_{1_0}}^{x_{1_f}}\left.\langle \mathbf{\mathcal{B}}^*\hat{\mathbf{q}}^*,\delta\hat{\mathbf{q}} \rangle\right\rvert_{0}^{x_{2_{max}}}\mathrm{d}x=\int_{x_{1_0}}^{x_{1_f}}\left.\left(V\overline{\hat{u}}^*\delta\hat{u}+(\overline{\hat{p}}+V\overline{\hat{v}}^*)\delta\hat{v}^*+V\overline{\hat{w}}^*\delta\hat{w}+\overline{\hat{v}}^*\delta\hat{p}\right)\right\rvert_{0}^{x_{2_{max}}}\mathrm{d}x_1=0
\end{equation}

\begin{equation}
\int_{x_{1_0}}^{x_{1_f}}\left.\langle \mathbf{\mathcal{C}}^*\hat{\mathbf{q}}^*,\delta\hat{\mathbf{q}}_y \rangle\right\rvert_{0}^{x_{2_{max}}}\mathrm{d}x_1=\int_{x_{1_0}}^{x_{1_f}}\left.\left(-\overline{\hat{u}}^*\delta\hat{u}_{x_2}-\overline{\hat{v}}^*\delta\hat{v}_{x_2}-\overline{\hat{w}}^*\delta\hat{w}_{x_2}\right)\right\rvert_{0}^{x_{2_{max}}}\mathrm{d}x_1=0
\end{equation}

\begin{equation}
\int_{x_{1_0}}^{x_{1_f}}\left.\langle \mathbf{\mathcal{C}}^*\hat{\mathbf{q}}^*_{x_2},\delta\hat{\mathbf{q}} \rangle\right\rvert_{0}^{x_{2_{max}}}\mathrm{d}x_1=\int_{x_{1_0}}^{x_{1_f}}\left.\left(-\overline{\hat{u}}^*_{x_2}\delta\hat{u}-\overline{\hat{v}}^*_{x_2}\delta\hat{v}-\overline{\hat{w}}^*_{x_2}\delta\hat{w}\right)\right\rvert_{0}^{x_{2_{max}}}\mathrm{d}x_1=0
\end{equation}

The boundary conditions are then given as,

\begin{equation}
\left.\overline{\hat{u}}^*=\overline{\hat{v}}^*=\overline{\hat{w}}^*\right \rvert_{x_2=0}=0
\end{equation}

\noindent and

\begin{equation}
\left.\overline{\hat{u}}^*=\overline{\hat{v}}^*=\overline{\hat{w}}^* \right\rvert_{x_2=x_{2_{max}}}=0
\end{equation}

Finally, zeroing the fourth boundary term, which comes from the streamwise derivative, supplies the initial condition for the adjoint variables, set in the final point of the domain, which corresponds to the objective position:

\begin{equation}
\int_{0}^{\infty}\left.\left((\overline{\hat{p}}^*+U\overline{\hat{u}}^*)\delta \hat{u}+U\overline{\hat{v}}^*\delta \hat{v}+U\overline{\hat{w}}^*\delta\hat{w}\right)\right\rvert_{x_{1_0}}^{x_{1_f}}\mathrm{d}x_2=\int_0^\infty \left.(\overline{\hat{\mathbf{q}}}-\overline{\hat{\mathbf{q}}_{SPOD}})\right\rvert_{x_1=x_{1_f}}\delta \hat{\mathbf{q}} \mathrm{d}x_2
\end{equation}

\noindent such that, at $x_1=x_{1_f}$, we have:

\begin{equation}
\label{terminal1}
\hat{u}^*(x_1=x_{1_f})=\hat{u}(x_1=x_{1_f})-\hat{u}_{SPOD}(x_1=x_{1_f}),
\end{equation}

\begin{equation}
\label{terminal2}
\hat{v}^*(x_1=x_{1_f})=\hat{v}(x_1=x_{1_f})-\hat{v}_{SPOD}(x_1=x_{1_f}),
\end{equation}

\begin{equation}
\label{terminal3}
\hat{w}^*(x_1=x_{1_f})=\hat{w}(x_1=x_{1_f})-\hat{w}_{SPOD}(x_1=x_{1_f}),
\end{equation}

\noindent and $\hat{p}^*(x_1=x_{1_f})=0$. 

After setting all these terms to zero, the variation of $J$ remains with a single term;

\begin{equation}
\label{finalequation1}
\delta J = \langle \hat{\mathbf{q}}^*,\delta \hat{\mathbf{F}\rangle}=\int_{x_{1_0}}^{x_{1_f}} \int_0^{\infty} \overline{\hat{q}^*}(x_1,x_2)\delta \mathbf{\hat{F}} \mathrm{d}x_2 \mathrm{d}x_1
\end{equation}

Since no source terms are being considered on the wall, the variation of $J$ may be written in terms of the gradient of the objective with respect to the forcing term,

\begin{equation}
\label{finalequation2}
\delta J = \int_{x_{1_0}}^{x_{1_f}} \int_0^{\infty} \overline{\nabla}_{\mathbf{F}}E_f\delta\mathbf{\hat{F}} \mathrm{d}x_2 \mathrm{d}x_1,
\end{equation}

\noindent which implies that 

\begin{equation}
\label{finalequation3}
\nabla_{\mathbf{F}}E_f=\hat{\mathbf{q}}^*.
\end{equation}

Equation \eqref{finalequation3} states that the gradient of the parameter to be optimized with respect to the forcing is equal to the adjoint variable and it therefore permits to find the desired forcing by means of a gradient scheme:

\begin{equation}
\label{equationforupdateoftheforce}
\mathbf{\hat{F}}^{n+1}=\mathbf{\hat{F}}^{n}+\gamma\nabla_{\mathbf{F}}E_f\delta\mathbf{\hat{F}}^n.
\end{equation}

The modification of the cost function, equation \eqref{costfunction}, will define the two actuators considered in this paper. If one considers $\hat{\mathbf{q}}_{SPOD}=0$, the optimization procedure will obtain the highest energy and by setting $\gamma > 0$ - this is referred to as the optimal forcing. If $\hat{\mathbf{q}}_{SPOD}$ is taken as the SPOD of the actual field, induced by the free-stream turbulence, then we take $\gamma < 0$, corresponding to a minimization - which is referred to as the identified actuator, which will target the specific structure inside the boundary layer. Other than the value of $\lambda$, the only change between the two methods is the terminal condition for the adjoint, eqs. \eqref{terminal1} --- \eqref{terminal1}.

The algorithm for the power iterations using the adjoint can then be written as:

\begin{enumerate}
\item Start with a random force field and zero initial conditions and integrate the direct problem, equation \eqref{blequations1}, from the position of actuation until objective.
\item Take the initial condition for the adjoint problem from \eqref{terminal1} --- \eqref{terminal2} and integrate equation \ref{lkfjaslfjsald} backwards, from the position of objective until the position of actuation.
\item Update the forcing field using equation \eqref{equationforupdateoftheforce} and calculate the cost function \eqref{costfunction}. Evaluate the convergence and either repeat from the first step or terminate the method.
\end{enumerate}

It could be advantageous to work with an actuator that is concentrated in the streamwise direction. To obtain this result, a Gaussian mask of the type of

\begin{equation}
\label{gaussianmask}
M(x_1)=\exp(-(x_1-x_{1_0})^2/L_{x_1}^2)
\end{equation}

\noindent may be used to multiply the forcing in equation \eqref{equationforupdateoftheforce} at each iteration. This leads to a slower convergence of the algorithm, however it was not found to be prohibitive.

Finally, the spanwise spatial support of the forcings were chosen to be also in the form of a Gaussian, which is multiplied by the final result of the forcing. The values of $(\omega,\beta)$ were chosen in accordance with the most amplified structures in the flow.

\bibliographystyle{jfm}
\bibliography{biblio}

\end{document}